\documentclass[fleqn,usenatbib]{mnras}

\DeclareRobustCommand{\VAN}[3]{#2}
\let\VANthebibliography\thebibliography
\def\thebibliography{\DeclareRobustCommand{\VAN}[3]{##3}\VANthebibliography}

\usepackage{newtxtext,newtxmath}
\usepackage[T1]{fontenc}
\usepackage{tgbonum}

\usepackage{graphicx}	% Including figure files
\usepackage{amsmath}	% Advanced maths commands
\usepackage{dsfont}
\usepackage{aas_macros}

\usepackage{comment}
\usepackage{xcolor}
\usepackage{ulem}
\usepackage{cancel}
\usepackage{soul}

\renewcommand{\vec}[1]{\ensuremath{\boldsymbol{#1}}}
\newcommand{\dd}{\ensuremath{\text{d}}}
\newcommand{\mm}{\ensuremath{\text{m}}}

\newcommand{\tnsr}[1]
{\ensuremath{{\fontfamily{cmss}\selectfont \textbf{\text{#1}}}}}
\newcommand{\Dxq}{\ensuremath{{\fontfamily{cmss}\selectfont \textbf{\text{D}}}_{\vec{x} \vec{q}}}}
\newcommand{\Dpq}{\ensuremath{{\fontfamily{cmss}\selectfont \textbf{\text{D}}}_{\vec{p} \vec{q}}}}
\newcommand{\dDxq}{\ensuremath{{\fontfamily{cmss}\selectfont \dot{\textbf{\text{D}}}}_{\vec{x} \vec{q}}}}
\newcommand{\dDpq}{\ensuremath{{\fontfamily{cmss}\selectfont \dot{\textbf{\text{D}}}}_{\vec{p} \vec{q}}}}

\newcommand{\sk}{\text{sk}}
\newcommand{\tth}{\text{th}}
\newcommand{\g}{\text{g}}
\newcommand{\M}{\text{M}}

\definecolor{mygreen}{RGB}{20,138,6}
\definecolor{myblue}{RGB}{96, 130, 182} %{52,180,230}
\definecolor{mypurple}{RGB}{239, 130, 13}
\definecolor{myred}{RGB}{202, 69, 71} 
\definecolor{cyan}{RGB}{117, 251, 253}
\definecolor{orange}{RGB}{242, 169, 59}
\definecolor{blue}{RGB}{0, 0, 245}
\definecolor{red}{RGB}{234, 51, 35}
\definecolor{gray}{RGB}{180, 195, 219}

\title[Excursion Sets with a "Perfect" Collapse Model]{Excursion Sets with a "Perfect" Collapse Model }

\author[A. M. Wislocka et al.]{
A. M. Wisłocka,$^{1}$\thanks{E-mail: agata.wislocka@univie.ac.at }
J. Stücker,$^{1,2}$
O. Hahn,$^{1}$
R. E. Angulo$^{2}$
\\
$^{1}$Institute for Astronomy, University of Vienna, Türkenschanzstraße 17, Vienna 1180, Austria\\
$^{2}$Donostia International Physics Center, Manuel Lardizabal Ibilbidea 4, Donostia 20018, Spain
}

\date{Accepted XXX. Received YYY; in original form ZZZ}

\pubyear{2025}

\begin{document}
\label{firstpage}
\pagerange{\pageref{firstpage}--\pageref{lastpage}}
\maketitle

\begin{abstract}
The $\Lambda$CDM model predicts structure formation across a vast mass range, from massive clusters ($\sim10^{15}\,\text{M}_\odot$) to Earth-mass micro-haloes ($\sim 10^{-6} \, \text{M}_\odot$), resolving which far exceeds the capabilities of current simulations. Excursion set models are the most efficient theoretical tool to disentangle this hierarchy in mass. We test the excursion set paradigm by combining smoothed initial density fields with a "perfect" collapse model -- $N$-body simulations. We find that a core excursion set assumption -- small-scale perturbations do not impact larger-scale collapse -- is approximately fulfilled but exhibits small quantitative violations dependent on the smoothing filter. For a sharp $k-$space cut-off $\sim 20\%$ of mass elements revert collapse as the smoothing scale decreases, while only $3.5\%$ do for a Gaussian and $5\%$ for a top-hat. Further, we test the simple deterministic mass-mapping $M \propto R^3$ (first-crossing scale to halo mass) relation. We find that particles that are first accreted into haloes at the same smoothing scale may end up in haloes of significantly different masses, with a scatter of 0.4-0.8 dex. We also demonstrate that the proportionally constant of this relation should be considered as a degree of freedom. Finally, we measure the mass fraction in different structure morphologies (voids, pancakes, filaments and haloes) as a function of filter scale. Typical particles appear to be part of a large-scale pancake, a smaller-scale filament and a notably smaller halo. We conclude that validating predictions of excursion set models on a particle-by-particle basis against simulations may enhance their realism.
\end{abstract}


%%%%%%%%%%%%%%%%% BODY OF PAPER %%%%%%%%%%%%%%%%%%

\section{Introduction}

A central prediction of the $\Lambda$CDM standard model of cosmology is that the density field of the early Universe has significant perturbations on a broad range of scales. These perturbations eventually collapse into dense dark matter haloes, with masses ranging from as large as $10^{15}\,\M_\odot$ (solar masses) to as small as a few $\M_{\earth}$ (Earth masses) or even smaller, for WIMP or QCD axion dark matter candidates.

The observed distribution of galaxies and clusters is closely linked to the formation of haloes with masses $M \gg 10^{9} \, \M_{\odot}$, making precise models of the halo mass function essential for reliably interpreting observations. Further, some observational probes may depend on the presence of much smaller mass objects. For example, perturbations in the arcs of strong gravitational lenses may indicate the presence of haloes of masses $M \sim 10^7 \, \M_{\odot}$ and smaller \citep{koopmans_2005, Vegetti_2009, 2015McKean, 2020Hsueh, 2024Gilman}. Further, if dark matter is composed of $\sim 100$ GeV WIMPs with a significant self-annihilation cross-section, it may even be possible to probe the smallest mass haloes $M \sim \M_{\earth}$ \citep[also known as `prompt cusps',][]{2005Diemand, 2010Ishiama, 2012WIMPannihilationBringmann, 2017Angulo, delos_2023_prompt_cusps, 2024MNRAS.52710802O}. Such objects may contribute most of the observable self-annihilation signal and might be an important factor in the cosmic gamma-ray background \citep{delos_2023_annihilation, stuecker_2023_cusp_encounters, delos_2024_igrb}. 

Therefore, it is important to reliably predict the abundance and properties of such structures. While the formation of high-mass objects ($M \gtrsim 10^{8} \, \M_{\odot}$) can be accurately traced with $N$-body simulations, lower-mass structures are generally inaccessible in standard representative volumes due to resolution constraints. A notable exception are the VVV simulations, which resolve 20 orders of magnitude in mass scale through zoom-in simulations of highly under-dense regions \citep{wang_2020_vvv, zheng_2024}. They, however, do not represent the general statistics of our universe. Analytical approaches are ideally suited to bridging this vast range of scales, allowing for a deeper understanding of structure formation in $\Lambda$CDM beyond what is currently achievable with N-body simulations. 

Arguably, the most successful analytical approaches to predict the abundance of collapsed structures, or the halo mass function (HMF), are the excursion set formalism \citep[][]{bond1991excursion}, and peak theory \citep{bardeen1986statistics}. Excursion set theory emerged as an improvement to the seminal work of \citet{press_1974} on obtaining a HMF analytically. In this approach, the density field is smoothed with a kernel at a scale $R_s$, thus allowing one to trace relevant non-local properties of the environment as local properties of the smoothed field. In the smoothed density field, it is then possible to detect collapse that is associated with the scale $R_s$ with a simplified method -- for example, a spherical collapse model \citep{press_1974} or an ellipsoidal model \citep{bond_myers_1996}. Subsequently, decreasing the smoothing scale allows one to identify which volume elements collapse on a given scale and then make a prediction for the HMF \citep[e.g.][]{press_1974, bond1991excursion, Sheth&Tormen1999, sheth2001, Tinker2008, maggiore2010, 2016Diespali}. At its core, this method is conceptually very simple, yet it delivers surprisingly accurate HMFs even when implemented in its most basic form.

Beyond the HMF, the excursion set method has numerous other applications in cosmology, yielding detailed predictions for various statistical quantities of structure formation. These include: Conditional mass functions \citep[e.g.][]{2008Rubino-condMF, shen2006excursion, 2011DeSimone-condMFandFormationTimes, 2017Tramonte-condMF}, mass accretion, merger histories and formation times \citep[e.g.][]{bond1991excursion, 1993Lacey_Cole, 1999Somerville-merger, 2000Somerville-merger, 2010Neistein-condMFandMergerHist, 2011DeSimone-condMFandFormationTimes} and the mass-concentration relation \citep[e.g.][]{Ludlow2016, 2021Bohr}, among others.

To achieve agreement with these statistical quantities excursion set formalisms tend to employ additional degrees of freedom that often go beyond the original excursion set formulation by \citet{bond1991excursion} and \citet{press_1974}. These degrees of freedom include the choice of the smoothing function for mass-mapping \citep{Zentner2007excursions, maggiore2010, 2018Leo-smoothKfilter, Delos2024} or the first-crossings distribution function \citep{bond1991excursion, Zentner2007excursions, Maggiore_2010, Farahi2013, schneider2013halo, delos_2024_igrb}, implementations of stochastic barriers \citep[e.g.][]{2011Ma, Maggiore_2010, 2011Corasaniti, 2013Achitouv, 2014Achitouv}, direct modifications of the spherical collapse threshold \citep{1995Monaco, 1996Eke, 2001Jenkins, Delos2024}, accounting for three-dimensionality of collapse \citep{bond_myers_1996, 2001Chiueh, sheth2001, 2006Lee, shen2006excursion, sandvik2007, stucker2018median, musso_2024} and other approaches to environmental dependence \citep{1998Catelan,2000Taruya}. In particular, ellipsoidal collapse models are especially compelling, as they may additionally capture the anisotropy of collapse of fluid elements. This anisotropy, also observed in simulations, leads to the formation of pancakes, filaments and haloes \citep{sheth2001, shen2006excursion, sandvik2007, 2011Ludlow}. However, since an accurate calculation of the triaxial effects on the evolution of fluid elements is rather challenging, \citet{sheth2001} and \citet{2002Sheth&Tormen} proposed an approximate approach that captures the dynamics statistically, with a set of free parameters and a moving barrier. If the parameters are calibrated against simulations they allow to describe HMF to very high accuracy at the qualitative level. For this reason this approach has achieved significant popularity. Nonetheless, this approach remains an oversimplification of the complex halo formation dynamics, with significant quantitative discrepancies reported between simulations and this model \citep{Reed_2006, 2009Robertson}.

While more complex excursion set approaches offer valuable, high-accuracy descriptions of statistical quantities, it remains uncertain whether this improved accuracy stems from a more realistic physical model or simply from the increased number of degrees of freedom. For example, the additional parameters in approximate ellipsoidal collapse models might effectively compensate for other shortcomings in the approach. Contrary to the common view, \citet{Delos2024} recently argued that this improvement should not be attributed to a more physically accurate modeling of the ellipsoidal collapse. Similarly, in a series of publications, \citet{2018Lucie, 2019Lucie, lucie_smith_2024} concluded that accounting for the tidal effects in the ellipsoidal collapse models has minimal impact on the HMFs. At the same time, experiments like those of \citet{2024Dani}, do uphold the claim of the importance of accounting for the environmental tidal effects in developing improved analytical methods for halo formation. Determining the true source of improvement is in particular relevant when making predictions about aspects that have not been rigorously tested. Less explored areas include predictions related to the cosmic web \citep{shen2006excursion, sandvik2007}, extrapolations to scales that remain unresolved in simulations \citep{angulo_2010, stucker2018median, Liu_2024}  or particle-by-particle predictions of the evolution of individual mass elements \citep{bond1991excursion, 2002MonacoPinocchio1, 2002MonacoPinocchio2, sheth2001}.

Here, we propose a novel approach to reliably test the fundamental assumptions that are common in all excursion set formalisms. For this, we evaluate the performance of the excursion set approach if the collapse model -- that is typically described by a spherical or ellipsoidal collapse approximation -- is instead replaced by a fully non linear simulation with a structure identification algorithm. We will refer to this as an excursion set with a `perfect' collapse model. The resulting excursion set draws an upper limit to the degree of realism that may be achieved by excursion set models. Higher accuracy may only be achieved if model parameters compensate for the short-comings in the excursion set paradigm itself.

We would like to emphasise that this type of analysis only became possible due to the phase-space simulations that do not suffer from artificial fragmentation, combined with a tessellation-based structure finder, which allows the identification of the morphology (halos, sheets, etc.) of fluid elements \citep{Stuecker:2020sheetsim, stucker2022hmf}. These breakthroughs in computational cosmology enable us to directly test certain assumptions of the excursion set theory for the first time. In this regard, our work complements that of \citet{2009Robertson} in the ongoing effort to test the groundwork of the excursion set theory.

This article is organized as follows: In Section \ref{sec:excsettheory}, we summarize key ingredients that are common among all excursion set formalisms. In Section \ref{sec:2}, we describe the details of our simulations, how collapse along different axes is detected and how we can use simulations as a ``perfect'' collapse model. In Section \ref{sec:3}, we present our measurements and test how well different assumptions of excursion sets are fulfilled. Finally, in Section \ref{sec:discussion_conclusion}, we discuss the inherent limitations of the excursion set formalisms. We also discuss which aspects of the excursion set formalisms can be improved to enhance their realism.

\section{Excursion set models} 
\label{sec:excsettheory}

Excursion set formalisms aim to disentangle the hierarchy of structure formation through four core ingredients:
\begin{itemize}
    \item The smoothed linear density field is considered a function of the smoothing scale $R$ -- allowing the tracking of relevant non-local properties of an object's environment as local properties of the smoothed environment. 
    \item A collapse criterion distinguishes between fluid elements that are part of a gravitationally collapsed structure and those that are not. 
    \item It is assumed that structure formation can be represented through an excursion set across smoothing scales: i.e. a mass-element is part of a structure given by the largest smoothing scale at which the collapse criterion is fulfilled.
    \item To define mass functions, it is assumed that the smoothing scale of collapse can be mapped to a halo mass through a simple one-to-one deterministic relation.
\end{itemize}
In this section, we will briefly summarize these core concepts of excursion set formalisms and point out a few popular examples. Finally, we will motivate that a simulation may be used as a ``perfect'' collapse model to test the core ingredients of excursion sets. By ``perfect'', we mean that no simplifying assumptions are made -- such as approximating collapse as spherical or ellipsoidal with a simple external tide model, etc.

\subsection{Scale space}
\label{subsec:scale_space}
In excursion set formalisms, it is common to define a smoothed linear density field $\delta_R(\vec{q})$ -- where $\vec{q}$ are Lagrangian coordinates and $R$ is a smoothing scale -- by convolving with a filter function
\begin{align}
    \delta_{R} &= W_R \ast \delta
\end{align}
If one considers the space of Lagrangian coordinates plus the smoothing scale $(\vec{q}, R)$, one may speak of a ``scale-space''. The smoothed density field can be conveniently evaluated in Fourier space
\begin{align}
    \widetilde{\delta}_{R} &= \widetilde{W}_R \cdot \widetilde{\delta}
\end{align}
where a tilde designates a Fourier transformation. 

Here, we will consider a few common choices of the filter function as the sharp k-space
\begin{subequations}
\label{eq:W_kRs}
    \begin{align}
    \widetilde{W}_{\sk}(kR) &= \Theta(k - k_R) \,,  \\
    \intertext{with $k_R = R^{-1}$, the top-hat} 
    \widetilde{W}_{\tth}(kR) &= \frac{3 \, [ \text{Sin}(kR) - kR \, \text{Cos}(kR) ]}{(kR)^3} \\
    \intertext{and the Gaussian}
    \widetilde{W}_{\g}(kR) &= \text{exp} \left [- \frac{(kR)^2}{2} \right ] 
\end{align}
\end{subequations}
where $\theta$ is the Heavyside step-function. Beyond the linear density field at the scale of consideration, the variance is of particular relevance
\begin{equation}
        \label{eq:sigma_R}
        \sigma_R^2 = \frac{1}{2\pi^2} \int^{\infty}_{0} k^2 \, P(k) \, \widetilde{W}^2(kR) \, \dd k\,.
\end{equation}
We will briefly discuss some of the characteristics of the above window functions. For a more in-depth discussion, we refer the reader to \citet{bond1991excursion} and \citet{maggiore2010}. For the case of a sharp-$k$ filter, $\widetilde{W}_{\sk}$, $\delta_R$ performs a Markovian random walk as a function of $R$. This makes it the only filter for which purely analytic expressions for the HMFs have been obtained. Conversely, any other filter leads to correlated random walks, which makes a fully analytical treatment difficult. However, the top-hat function $\widetilde{W}_{\tth}$ seems to be the most natural choice when working with a spherical collapse model, and the mass associated with a collapsed Lagrangian patch is well-defined in real space. Similarly, The Gaussian window $\widetilde{W}_{\g}$ has the convenience of being well defined and localised in configuration space, and yields well-defined results on the statistics of peaks, contrary to the $\widetilde{W}_{\sk}$ and $\widetilde{W}_{\tth}$ functions \citep{bardeen1986statistics}.
In general, filter functions are chosen based on the goal of recovering some target statistic. For example, \citet{schneider2013halo} have demonstrated that, of these three filters, the sharp $k-$space filter is the only one to recover the asymptotic low-mass behaviour of HMFs in warm dark matter cosmologies with a cut-off in the power spectrum, as observed in simulations \citep{2013BensonWDM, schneider2013halo, 2015SchneiderWDM, 2018Leo-smoothKfilter}. In contrast, \citet{Delos2024} has shown that for linear power spectra without a cut-off, i.e. CDM cosmologies, the top-hat window function gives very accurate results for many scenarios when no approximations are used in the excursion set formalism.

Clearly, there is no unique optimal window function for all scenarios. Therefore, in this article, we will test how well it is justified to disentangle structure formation in scale space for all three window functions.

\subsection{The collapse criterion}
\label{subsec:coll_criterion}
Excursion set formalisms define a collapse criterion that distinguishes between fluid elements that are assumed to collapse and those that do not. While the primary focus of most formulations of the excursion set \citep{bond1991excursion} is  to distinguish between haloes and non-haloes, it is possible to consider more general collapse definitions that classify collapsed structures into voids, pancakes (or walls), filaments and haloes. To handle these structures systematically, we define a \textit{morphology rank} as a number $n$ that distinguishes structures as follows: 0 = \textit{void}, 1 = \textit{pancake}, 2 = \textit{filament} and 3 = \textit{halo}. We may understand $n$ as the number of axes of a volume element that have collapsed \citep[see, e.g.][]{Falck:2012origami}. This definition implicitly assumes that a volume element within a halo has previously undergone collapse into a filament and a pancake, representing another form of hierarchical structure formation.

We may then define a collapse criterion as a functional that maps a smoothed linear density field onto a morphology rank.
\begin{align}
    c : \delta_R(\vec{q} - \vec{q}_0) \rightarrow n \in \{0, 1, 2, 3\}
\end{align}
which may be evaluated separately for each location of interest $\vec{q}_0$. Note that this definition includes translation invariance and, as a functional, allows to incorporate aspects of the linear density field beyond the local density $\delta_{R} = \delta_{R}(\vec{q} - \vec{q}_0 = 0)$, such as the deformation tensor.

While we do not use analytical collapse models in this study, we still want to mention two important examples for illustration: The most commonly adopted collapse criterion is that of spherical collapse \citep{press_1974}, which assumes a volume element to be collapsed into a halo if the smoothed linear density exceeds a value of $\delta_c = 1.68$:

\begin{align}
    c_{\mathrm{sc}} = \begin{cases}
        3 \text{\quad if \quad} \delta_{R} \geq \delta_c \,, \\
        0 \text{\quad otherwise \,. \quad}
    \end{cases}
\end{align}
More sophisticated triaxial or ellipsoidal collapse models may distinguish between the collapse along each of the three axes, based on the deformation tensor $d_{R,ij} = (\partial_i \partial_j / \nabla^2) \delta_R$:
\begin{align}
    c_{\mathrm{ec}} = \begin{cases}
        3 \text{\quad if \quad} \delta_{R} \geq f_{\mathrm{ec,h}}(e_R, p_R) \,, \\
        2 \text{\quad else if \quad} \delta_{R} \geq f_{\mathrm{ec,f}}(e_R, p_R) \,, \\
        1 \text{\quad else if \quad} \delta_{R} \geq f_{\mathrm{ec,p}}(e_R, p_R) \,,\\
        0 \text{\quad otherwise \,, \quad}
    \end{cases}
\end{align}
with three separate barriers for pancake, filament and halo formation that depend on the ellipticity $e_R$ and the prolateness $p_R$ of the deformation tensor $d_{R,ij}$, which are defined through its eigenvalues $\lambda_1 \geq \lambda_2 \geq \lambda_3$ as $e_R = (\lambda_1 - \lambda_3)/2(\lambda_1 + \lambda_2 + \lambda_3)$ and $p_R = (\lambda_1 + \lambda_3 - 2 \lambda_2)/2(\lambda_1 + \lambda_2 + \lambda_3)$. Examples of triaxial models are the tide-free ellipsoidal collapse model of \citet{white_1979}, the ellipsoidal collapse model with tides by \citet{bond_myers_1996}, the ellipsoidal approximation (only for halo formation) from \citet{sheth2001}, the (shape-free) triaxial collapse model in \citet{stucker2018median} or the Lagrangian perturbation theory expansion of the triple sine-wave considered by \citet{rampf_2023}.

In principle, collapse models could incorporate even higher-order aspects of the linear field into the collapse criterion. However, it is unclear a priori how fruitful this approach would be, as other aspects of an excursion set model, such as mass assignment \citep{Sheth&Tormen1999}, might fail independently of the degree of realism employed in the collapse model.

\subsection{The excursion set assumption} 
\label{sec:excursionsetassumption}
Once a collapse criterion has been defined, it is possible to test whether this criterion is satisfied when applying filters to the linear field at different smoothing scales $R$. We will write ``$c_R$'' as a short-hand for the collapse criterion evaluated on the density field at smoothing scale $R$. Generally, we may encounter different classifications at different length scales. For example, according to one criterion, it may be, that a particle is considered part of a halo $c_{R_1} = 3$ on some scale $R_1$, but outside of a halo $c_{R_2} < 3$  on a larger scale $R_2 \geq R_1$. It is the core assumption of the excursion set formalisms that a particle has the morphology rank $n_R$, if it was determined to have been collapsed to a structure of this rank on \textit{any} larger scale:
\begin{align}
    n_R &= \sup \{ c_{R'}  \,|\, R' \geq R\}. \label{eqn:excursionset}
\end{align}

In other words, this would mean that the particle is predicted to be part of a large-scale halo and that the density perturbations on smaller scales are irrelevant to the collapse question. 

Note that here we have defined an excursion set on the \textit{morphology rank}, whereas the original spherical collapse formulation by \citet{bond1991excursion} was defined as an excursion set on the \textit{smoothed linear density field} $\delta_R$, that is then compared with the spherical collapse barrier $\delta_c$ (giving a classification $n_R$). In the case of spherical collapse, these two definitions are equivalent. However, defining the excursion set on the morphology rank provides a clear extension for more general collapse models.

Each collapsing axis has an associated length scale at which the corresponding collapse criterion is first fulfilled. In particular, we can define the Lagrangian length scales of pancakes, filaments and haloes:
\begin{subequations}
\label{eq:sim_ES_def}
    \begin{align}
    R_{\mathrm{pancake}} &= \sup \{ R \, | \, c_R \geq 1 \} \label{eqn:rpancake} \\
    R_{\mathrm{filament}} &= \sup \{ R \, | \, c_R \geq 2 \} \label{eqn:rfilament}\\
    R_{\mathrm{halo}} &= \sup \{ R \, | \, c_R = 3 \} \label{eqn:rhalo}
\end{align}
\end{subequations}
If the corresponding sets are empty, indicating that a particle has not collapsed at any length scale, these radii are considered undefined. Note that these are Lagrangian scales and are therefore related to mass scales, but they should not be mistaken for the Eulerian extent of these structures, as the two can differ significantly. Further, note that the definitions imply that strictly $R_{\mathrm{pancake}} \geq R_{\mathrm{filament}} \geq R_{\mathrm{halo}}$. That is, every halo is embedded in a larger-scale filament, which in turn is part of an even larger-scale pancake. At sufficiently large scales, these structures are not collapsed at all. The latter aspect is, for instance, utilised in the EFTofLSS (effective field theory of large-scale structure) models \citep[cf.][]{Baumann:2012,Carrasco:2012}, where all highly non-linear scales are integrated out in order to guarantee the applicability of perturbative techniques on large-enough scales.

\subsection{The mass map}
\label{subsec:mass_map}

The previous definitions already allow for the association of each fluid element with a scale of collapse (for a given collapse model). However, to obtain predictions, e.g. of the halo mass function, it is additionally necessary to relate the collapse scale to a mass scale. Therefore, all excursion set formalisms (that we know of, with the exception of \citealt{HahnParanjape:2014}) assume that \textit{if a fluid element first satisfies the halo collapse criterion at a smoothing scale $R$, then it is part of a halo with mass}
\begin{align}
    M_{\mathrm{halo}} = M(R_{\mathrm{halo}}) \,.
\end{align}
It is worth stressing that this relation is assumed to be \textit{deterministic} and that all fluid elements that fullfil the collapse criterion at the same smoothing scale become part of a halo of the same mass. This is a point that we will investigate in more detail later.
Generally, all implementations assume that the mass map should scale as $M(R) \propto R^3$. However, there is no universally accepted method for defining the proportionality constant of the relation, except in the case of the top-hat filter. Assuming a unit filter normalization, $\int W_R = 1$ \citep{maggiore2010}, one has
\begin{align}
    \label{eq:mass} 
        \frac{M(R)}{\rho_{m,0} \, R^3} = 
    \begin{cases}
         \;\; 6 \pi^2  & \text{sharp $k$-space}, \\
         \;\; 4 \pi / 3 & \text{top-hat}, \\
         \;\; (2 \pi)^{3/2}   & \text{Gaussian} ,
    \end{cases}
\end{align}
where $\rho_{\mm,0}$ is the mean matter density at $z=0$. We will also assume (\ref{eq:mass}) as the baseline for the mass-map in our analysis.

\subsection{Excursion sets with a perfect collapse model}

In general, excursion set formalisms assume simplistic collapse models to describe aspects of the evolution of the fully nonlinear density field, such as the halo mass function. Inaccuracies in the description of the modeled quantities have often been attributed to shortcomings in the realism of collapse models. For instance, it is widely accepted that an ellipsoidal collapse model instead of spherical collapse leads to an improved description of the halo mass function \citep{sheth2001}.

In the next section, we will discuss how adopting a simulation as a collapse model is possible. This allows us to test how well excursion sets perform if the degree of realism of the collapse model is maximized. For simplicity, we call this a "perfect" collapse model. This approach allows us to distinguish between inaccuracies arising from the realism of the collapse model and the limitations arising from the mathematical structure of the formalism outlined above.

\section{Simulation as a collapse model}
\label{sec:2}

In this section, we describe how a fully non-linear simulation can serve as a collapse model within an excursion set formalism. In the mathematical language we adopted in Section \ref{sec:excsettheory}, any functional that predicts morphology rank for a given linear density field qualifies as a collapse model. To use a simulation as a collapse model, we may, therefore, smooth the linear density field at a scale $R$, run a simulation with such modified initial conditions and subsequently determine collapsed regions in the simulated fields to define a morphology rank for each particle. The excursion set may then be obtained by simulations with initial conditions smoothed on many scales.
 
In Sections \ref{subsec: EoM} through \ref{subsect:axis_collapse} we will describe how we infer the morphology rank for particles in a simulation and in Section \ref{subsec:excsims}, how we set up a large number of simulations to mimic excursion sets.

\subsection{Equations of motion}
\label{subsec: EoM} 
According to the current paradigm, DM is collisionless, cold and made up of purely gravitationally (or possibly also weakly) interacting particles \citep{Bertone2018DM}. The equations of motion (EoM) for such non-relativistic, collisionless and self-gravitating matter in terms of a (co-moving) position $\vec{x}(\vec{q},t)$ and canonically conjugate momentum $\vec{p}(\vec{q},t)$ in terms of cosmic time $t$ and indexed by a Lagrangian coordinate $\vec{q}$ (defined as $\vec{q}=\vec{x}(\vec{q},t=0)$) are given by the following ordinary differential equations \citep[e.g.][]{Peebles:1981} 
\begin{subequations}
\label{eq:sim_eqn}
   \begin{align}
   &\dot{\vec{x}}(\vec{q},t) = a^{-2}(t)  \, \vec{p}(\vec{q},t) \\ 
   &\dot{\vec{p}}(\vec{q},t) = - a^{-1}(t) \, \vec{\nabla}_{\vec{x}} \phi(\vec{x},t) \\
   &\vec{\nabla}^2_x \phi(\vec{x},t) = 4 \pi \, G \rho_{{\rm m},0} \, \delta (\vec{x},t) \label{eq:Poissin_p} \,,
\end{align} 
\end{subequations}
where a dot denotes a derivative w.r.t. cosmic time, $a(t)$ is the scale factor, $\phi(\vec{x},t)$ is the peculiar potential, $\delta(\vec{x},t) = \frac{\rho(\vec{x},t) - \rho_{\mm}(t)} {\rho_{\mm}(t)}$ is the matter density contrast, $\rho_m(t)$ is the time dependent background matter density. These equations of motion are solved in the \textsc{L-Gadget3} $N$-body code that we use for our numerical simulations \citep{springel2005cosmological,angulo2012scaling}. The perfectly cold limit is already implicit in our notation, in that we think of the fluid as dispersionless in Lagrangian space, i.e. initially, the momentum distribution at location $\vec{q}$ is single-valued. 

In this study, we deliberately exclude any baryonic effects, as the primary objective is to evaluate the assumptions of the excursion set theory only in the context of DM structure formation. Moreover, incorporating baryonic effects is unlikely to significantly influence the simulation outcomes related to the aspects of our investigation. For the smallest DM haloes ($\M \sim \M_{\earth}$) baryons tend to stream out of the halo potentials, leaving these structures unaffected.

\subsection{The distortion of fluid elements}
\label{subsec: Dxq}

Given its cold nature, i.e. the lack of significant primordial velocity dispersion, the DM distribution is reduced to a 3-dimensional sub-manifold (parameterised by the Lagrangian coordinate $\boldsymbol{q}$) embedded in a 6-dimensional phase-space; this sub-manifold is often referred to as the DM \textit{phase-space sheet} \citep{Abel:2012sheet, Shandarin:2012sheettesselations}.

The EoMs \eqref{eq:sim_eqn} imply a deformation (or distortion) of each fluid element comprising this sheet, which is described by the Jacobian matrix of the Lagrangian map 
\begin{align}
   \Dxq (\vec{q},t) &:= \frac{\dd\vec{x} (\vec{q},t)}{\dd\vec{q}} \,.
\end{align}
Note that in general, $\Dxq$ is not a symmetric tensor. However, the antisymmetric part only represents an axis rotation, hence it only becomes important in a highly non-linear setting, i.e. post shell-crossing.

As the phase-space sheet evolves, it gradually creases under the influence of gravity. In this description, the distortion tensor is directly related to the relative density change of an infinitesimal volume of the DM sheet (i.e. fluid element) as the sheet changes shape. As a direct consequence of mass conservation, the fluid particle density is given by the so-called \textit{stream} density
\begin{align}
    \label{eq:GDE_dens_contrast}
    \rho_s(\vec{q}, t) =  \frac{\rho_m(t)}{| \det \Dxq (\vec{q},t) |} \,.
\end{align}

At early times, there is a one-to-one mapping between Lagrangian and Eulerian space. At later times, the mapping will generally become multi-valued (i.e. `multi-stream'), and multiple Lagrangian fluid elements will map to the same Eulerian coordinate. Specifically, if one axis collapses along a given direction $\vec{\xi}$, then $\Dxq\cdot\vec{\xi}$ flips its sign at that instant, invalidating the bijective property of the Lagrangian mapping. In other words, the fluid element shell-crosses along $\vec{\xi}$. This, in turn, results in a singularity in $\rho_s$, as the determinant will be zero at that moment. The distortion tensor thus entails all (infinitesimal) information about the three-dimensional deformation and possible triaxial collapse of each fluid element.

In the literature, two distinct methods have been discussed that can yield the evolution of $\Dxq$ in a fully non-linear cosmological setting, with general random initial conditions. These are the GDE approach \citep[geodesic evolution equation;][]{vogelsberger2008fine} and the Lagrangian sheet tessellation approach \citep{Abel:2012sheet,Shandarin:2012sheettesselations,Hahn:2016adaptivesheet,Sousbie:2016coldice,Stuecker:2020sheetsim}. In this paper we will focus on the former.

Unlike a standard $N$-body simulation, which follows only particle positions and momenta, the advantage of the simulation we employ for this study is that it follows the per-particle temporal evolution of the distortion tensor with the GDE,
\begin{subequations}
\label{eq:GDE}
   \begin{align}
   \dDxq &= \frac{\dd}{\dd t} \left ( \frac{\dd\vec{x}}{\dd\vec{q}} \right ) = a^{-2} \; \Dpq \label{eq:chi_dot} \\ 
   \dDpq &= \frac{\dd}{\dd t} \left ( \frac{\dd\vec{p}}{\dd\vec{q}} \right ) = a^{-1} \; \tnsr{T}\cdot \, \Dxq \,, \label{eq:eta_dot}
\end{align} 
\end{subequations}
where $\Dpq := \frac{\dd\vec{p}}{\dd\vec{q}}$ and $\, T_{i j} = - \, \partial_i \partial_j \phi \,$  is the tidal tensor. For an in-depth description of the distortion tensor and the GDE, refer to \citet{vogelsberger2008fine} and \cite{vogelsberger2011streams}.

Our modified version of \textsc{L-Gadget3} \citep[cf.][for details]{Stuecker:2020sheetsim} uses a standard cosmic-time second-order leapfrog algorithm to integrate eqs.~\eqref{eq:GDE} in time. This enables us to follow the volume distortion over time and infer from it the cosmic web morphology class of all simulation particles. 

\begin{figure}
	\includegraphics[width=\columnwidth]{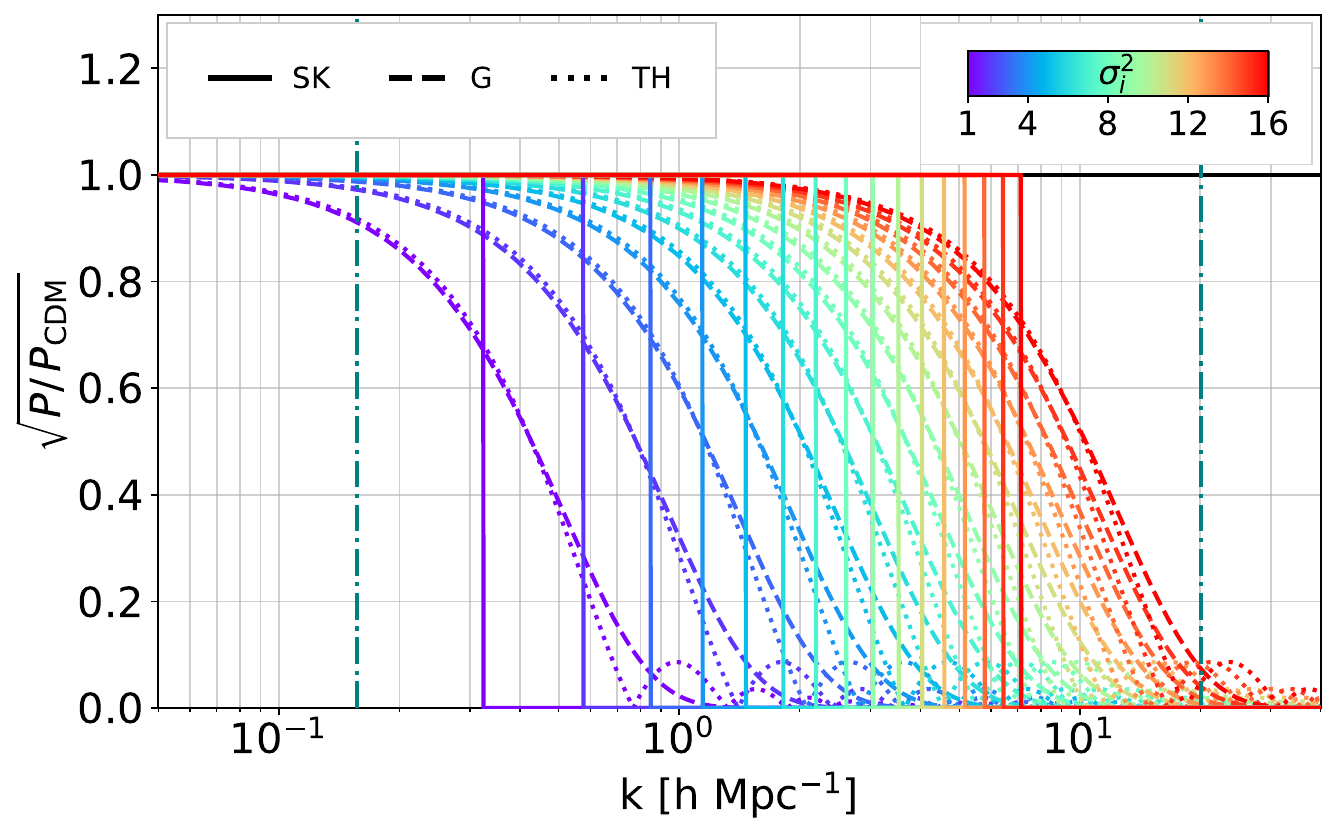}
    \includegraphics[width=\columnwidth]{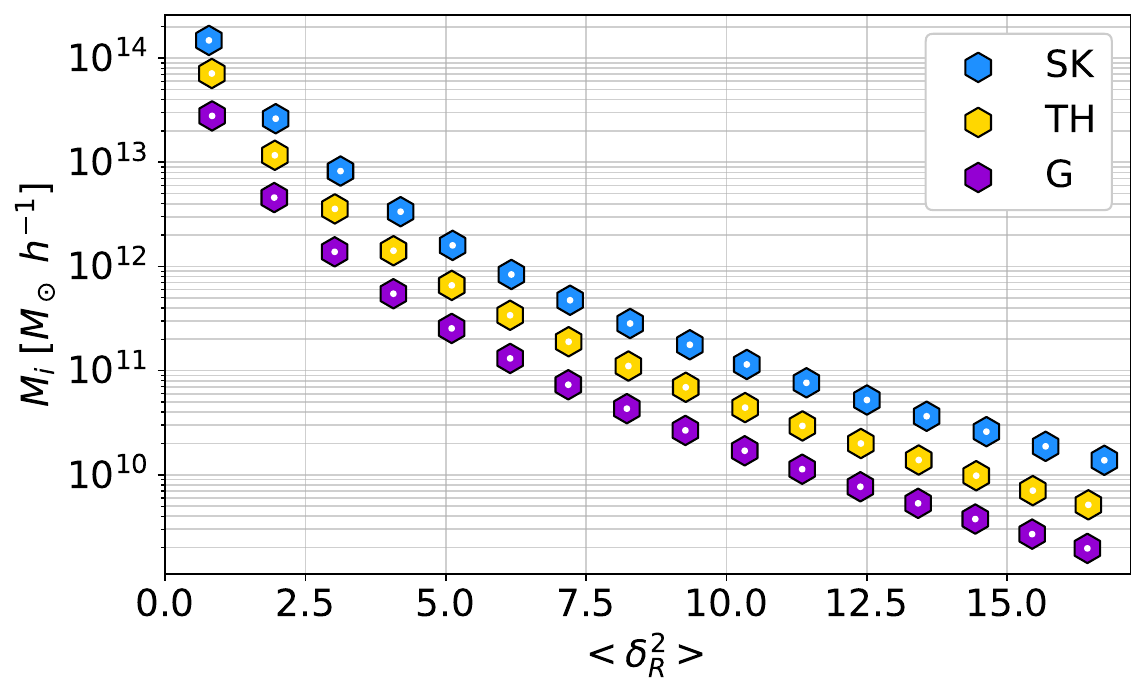}
    \caption{\textbf{Top:} The initial power spectra ($z=49$) generated with the $\widetilde{W}_{\sk}$ (continuous), $\widetilde{W}_{\tth}$ (dotted) and $\widetilde{W}_{\g}$ (dashed) smoothing on 16 different scales, normalized to the CDM spectrum (black). Each spectrum uses a sharp cut in Fourier space to approximately achieve the target variance $\sigma_i^2$. The two teal dash-dotted lines indicate the min. (left) and max. (right) $k-$modes of our simulation volume. \textbf{Bottom:} The characteristic mass associated with each $k_R$ scale as a function of the measured variance of the simulation for the sharp $k$-space (blue), top-hat (yellow) and Gaussian (purple) filters.}
    \label{fig:P_cuts}
\end{figure}

\subsection{Detecting axis collapse}
\label{subsect:axis_collapse}
There is an evident link between axis collapse and the morphological characteristics - voids, pancakes, filaments and haloes - of the cosmic web, within which simulation particles are embedded \citep{Zeldovich:1970, Bond:1996cosmicweb, Falck:2012origami,Shandarin:2017flipflop, Ramachandra:2017web, Stuecker:2020sheetsim}. In essence, by tracing particles from their initial positions, one can tell which structures they end up in, while at the same time recording the correlation between their position and the number of collapsed axes. This tracing of particle morphology/collapse constitutes a substantial aspect of the excursion set theory. Since we will frequently reference these structural components of the cosmic web throughout the text, this subsection details the methodology adopted by \citet{Stuecker:2020sheetsim} for classifying fluid elements into the four morphological categories within the simulation code utilized in this study.

Following $\Dxq$ over time provides insight into the infinitesimal volume deformation and rotation, eqs.~\eqref{eq:GDE}, at discrete locations centered on the $N$-body particle positions \citep{Stuecker:2020sheetsim}. The non-symmetric aspects of the tensor describe rotations, while the symmetric aspects represent the volume deformation along the three dimensions. In the absence of rotations, $\Dxq$ is symmetric and, therefore, can be diagonalised with real eigenvalues, in which case the diagonal elements are simply the eigenvalues that, in turn, describe the stretching of the axes in the principle axis frame (i.e. the initial orientation of the axes). This rotation-less scenario occurs as long as the evolution remains linear. However, the growth of gravitational instability in time leads to non-linear evolution - activating the non-symmetric components of $\Dxq$  - and finally collapse along one, two or all three of the axes. As the evolution of the sheet progresses from linear to non-linear, the orientation of fluid elements inevitably changes, as the DM sheet begins to twist and eventually folds. This folding marks the moment of shell-crossing, and a fluid element at its location is turned inside out.

This sign flip, or rotation, can be traced by comparing the initial (Lagrangian) and final (Eulerian) orientation of a fluid element. The simulations of \citet{Stuecker:2020sheetsim}, that we employ, use the degree of rotation as a classification scheme for collapse. Specifically, they consider the singular-value decomposition (SVD) of $\Dxq(t)$: 
\begin{equation}
    \Dxq = \tnsr{USV}^\top \,,
\end{equation}
where \tnsr{S} is a diagonal matrix with singular values $s_i$ along the diagonal, and \tnsr{U} and \tnsr{V} are orthogonal matrices, where \tnsr{V} and \tnsr{U} impart the orientation to a volume element in the Lagrangian and the Eulerian configurations, respectively. Simply put, the SVD can be thought of as a generalization of the eigenvalue decomposition to non-symmetric matrices, so that here, the $s_i$ quantifies the degree of stretching of the principal axes. They define the rotation angles
\begin{equation}
    \alpha_i = \cos ^{-1}(\vec{v}_i \cdot \vec{u}_i) \,,
\end{equation} where $\vec{v}_i$ and $\vec{u}_i$ are the column vectors of \tnsr{V} and \tnsr{U}. The $\alpha_i$ indicate the degree of rotation of those axes with respect to their initial orientation. \citet{Stuecker:2020sheetsim} have found that particles that are part of different types of structures (voids, pancakes, filaments and haloes) show very different evolution of $\alpha_i$. Along the axes where a structure is not collapsed $\alpha_i$ is very close to zero, whereas its value along collapsed dimensions tends to vary rapidly between 0 and $\pi$.

For this reason, the simulation traces the time-maximum angle 
\begin{align}
    \alpha_{\mathrm{max}, i}(t) &= \sup \{\alpha_{i}(t')  \,\, | \,\, t' \leq t \}
\end{align}
to determine axis collapse. If the maximum angle of an axis exceeds 
a threshold of $\nu = \pi/4$ this axis is classified as collapsed. Following our notation introduced in Section \ref{subsec:coll_criterion} we can define the number of collapsed axes identified by the simulation for each fluid element:
\begin{align}
    n = \sum_{i=1}^3 \Theta(\alpha_{\mathrm{max},i} - \nu) \,,
\end{align}
where $\Theta$ is the Heaviside step function. \citet{Stuecker:2020sheetsim} observed that $n$ is relatively insensitive to the precise value of $\nu$, as $\alpha_{\mathrm{max},i}$ is typically either close to $0$ for axes that have not collapsed or close to $\pi$ for those that have.
Further, they have found that this classification is well able to distinguish between different morphologies and we will assume that $n$ corresponds to the morphology rank so that structures can be distinguished based on $n$ as described in Section \ref{subsec:coll_criterion} ( 0 = \textit{void}, 1 = \textit{pancake} etc.) We refer the reader to \citet{Stuecker:2020sheetsim} for further details.

Note that in our setup, $n$ may refer either to the target morphology or to a collapse classification criterion (denoted as ``$c$'' earlier) depending on the context. We will clarify this in the corresponding places in the text.

\subsection{An Excursion Set of Simulations}
\label{subsec:excsims}
For the purpose of our investigation we build a simulation set that mimics an excursion set. We do this by evaluating the above described morphology classification for a variety of linear density fields that are smoothed on different scales by simulating each of those initial conditions.

We initialise each of our simulations at $z=49$ ($a=0.02$) and we set the cosmological parameters to $\Omega_{\mathrm{m}} = 0.3051$, $\Omega_{\mathrm{\Lambda}} = 0.6948$, $h=0.676$, $n_s = 0.961$ and $\sigma_8 = 0.8154$. For all simulations, we use a box size of $L = 40\,h^{-1}\text{Mpc}$ with $N = 256^3$ particles to trace the dark matter sheet and a larger number of up to $N = 512^3$ released N-body particles. Since, in this work, we are not concerned with the accuracy of the internal structure of haloes and our interest lies simply in identifying the collapse times of various structures, we calculate forces only with a pure particle mesh approach with $N_{\mathrm{pm}} = 512^3$ grid elements leading to a force resolution of $\epsilon = 70 \,\mathrm{kpc}$.

For our investigation, we choose a set of 16 different smoothing scales $R_i$ for each of the three window functions discussed in Section \ref{subsec:scale_space}, so that $\sigma_{i}^2 := \sigma^2_{R_{i}}$
takes on a range of pre-specified values. We modified the MUSIC \citep{hahn2011multi} program to use eq.~\eqref{eq:sigma_R} together with a $\Lambda$CDM power spectrum $P(k)$ to output fields truncated with the appropriate cut-off, as shown in the top panel of Figure \ref{fig:P_cuts}. We select the target variances as $\sigma^2_i = i$, where $i=[1,...,16]$, and throughout this paper, we refer to each simulation by its corresponding target variance. Note that the measured variance of the initial conditions $\langle \delta_R^2 \rangle$ often turned out slightly different to the target variance due to cosmic variance. 
We show the relation between the variance of the density field and the characteristic mass scale as in equation \eqref{eq:mass} in the bottom panel of Figure \ref{fig:P_cuts}.

\begin{figure}
    \centering
    \includegraphics[width=\linewidth]{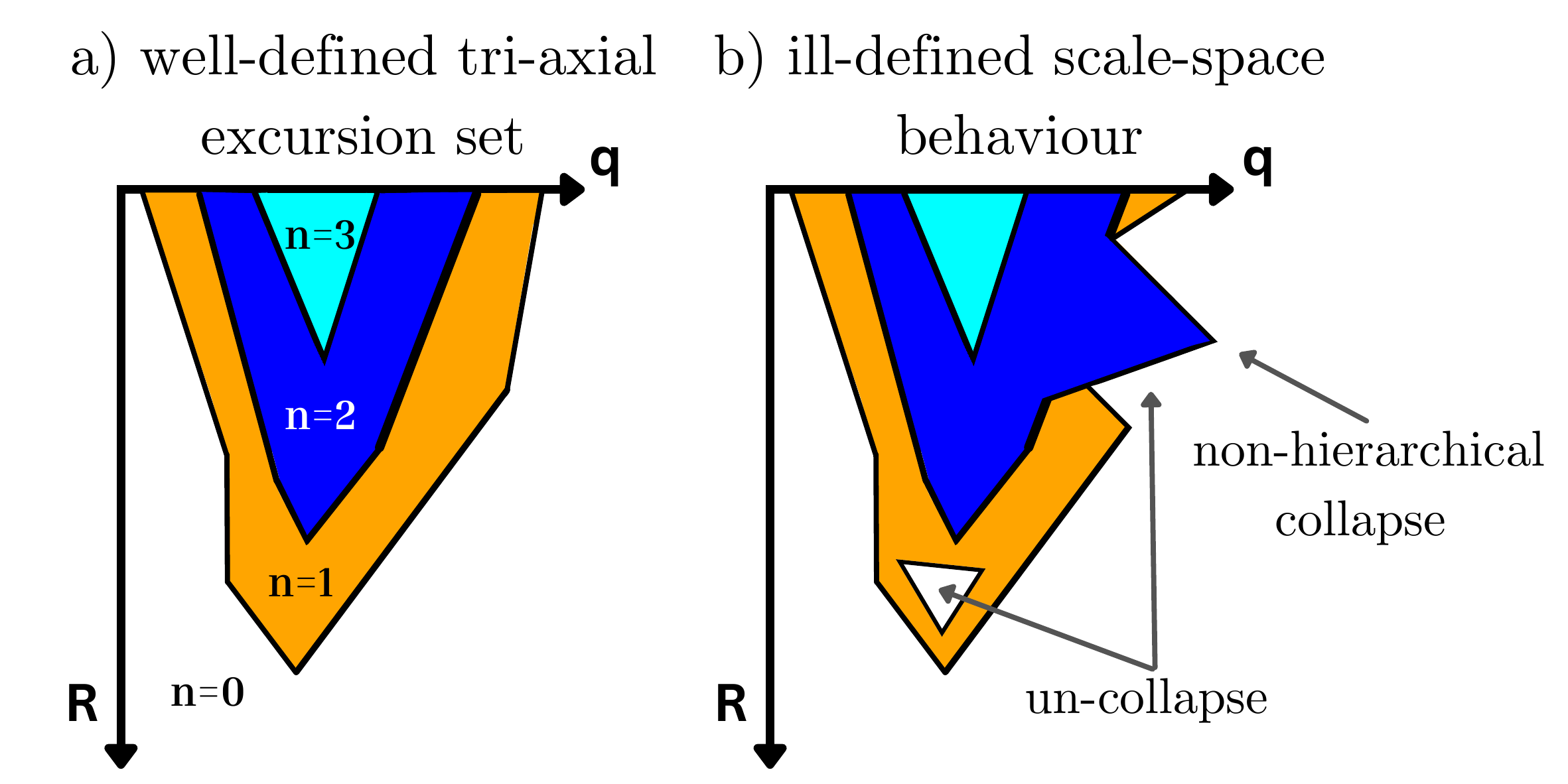}
    \caption{An illustration of an evolving Lagrangian field in the excursion set framework. \textbf{(a)} A well-defined case, where if one was to draw a vertical line starting from any point $\vec{q}$, the $n_R$ would decrease hierarchically with increasing $R$. The colour coding associated with the structure rank is white for $n_R = 0$ \textit{(void)}, orange for $n_R = 1$ \textit{(pancake)}, blue for $n_R = 2$ \textit{(filament)}, and cyan for $n_R = 3$ \textit{(halo)}. \textbf{(b)} An ill-defined example, where $n_R$ no longer decreases in a stepwise manner (e.g. a fluid particle to the right of the plot un-collapses from a pancake to a void and then collapses straight into a filament). }
    \label{fig:scale-space-excursion}
\end{figure}

\section{Emulation Excursion Sets with Simulations}
\label{sec:3}
Emulating excursion sets with our suits of simulations allows us to test basic assumptions of the excursion set theory independent of the assumed collapse model.

\subsection{The excursion set assumption}
\label{subsec:4.1}

The assumption that structure formation can be represented as an excursion set that progresses from large to small scales (see Section \ref{sec:excursionsetassumption}) is a powerful tool for understanding the hierarchical emergence of structures. In particular, it simplifies the picture by decoupling the formation of large-scale structures from density perturbations on smaller length scales. We may phrase the necessary assumption: 
\noindent \textit{If a volume element appears to collapse in the linear field smoothed on some scale, then it will become part of a structure of that length scale -- independently of the perturbations on smaller scales.} 

This assumption may be stated in condensed form as
\begin{align}
    n_{R_2} \geq n_{R_1}   \mathrm{\quad if \quad} R_2 \leq R_1 \label{eqn:excursion_inequality}
\end{align}
and is a direct consequence of eqn. \eqref{eqn:excursionset} if $c_R = n_R$ is defined through a perfect collapse model. In particular, it implies that no volume elements \textit{``un-collapse''} when smoothing scales are decreased. For example, let us consider a Lagrangian volume element that collapses in a simulation with some large smoothing scale $R_1$ into a filament ($n_{R_1} = 2$). In that case, it should also collapse into a filament (or a halo) in any simulation with a smaller smoothing scale $R_2 < R_1$. However, it should not become part of a lower rank structure ($n_{R_2} < 2$) like a pancake or void. 

\begin{figure}
    \includegraphics[width=\columnwidth]{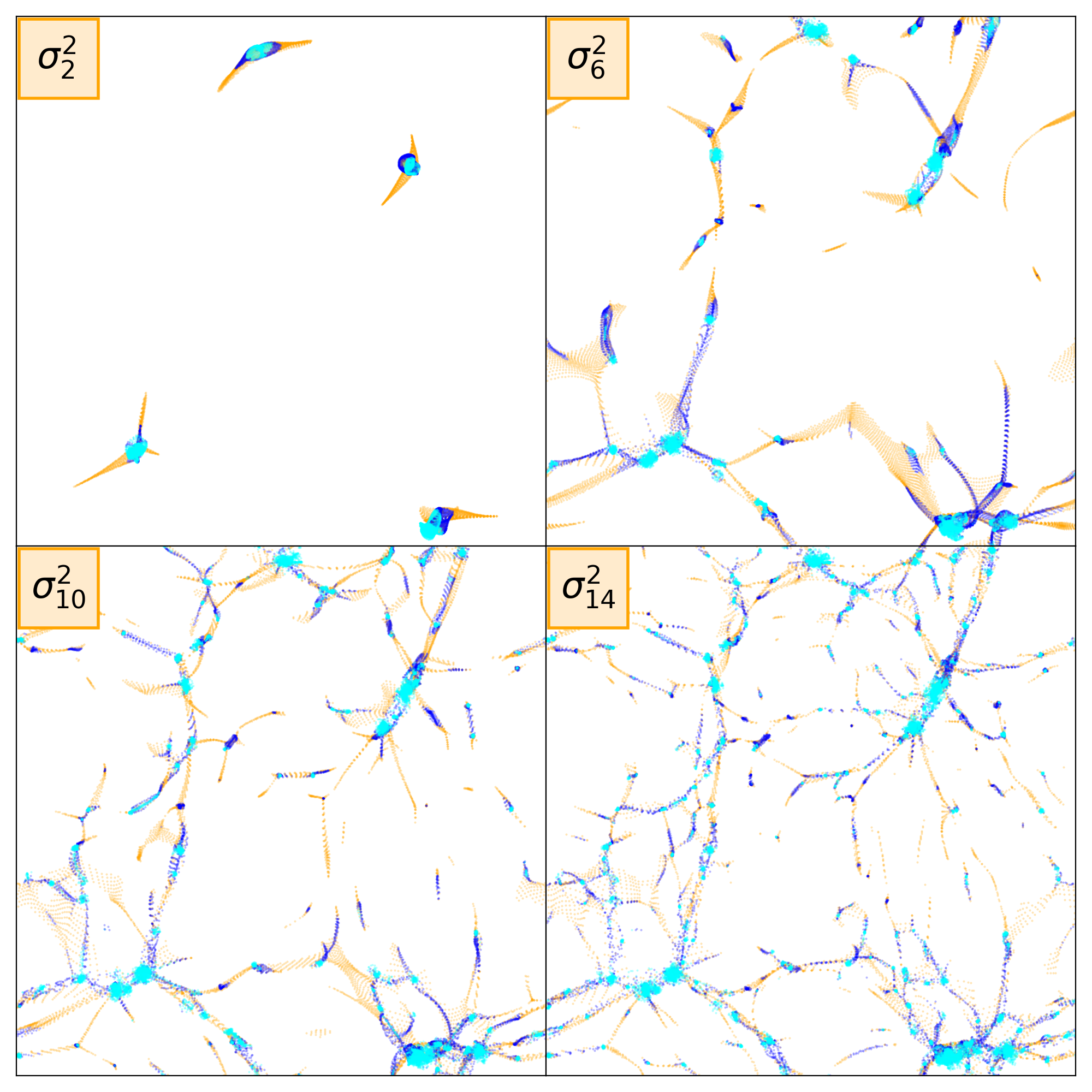}
    \caption{Thin slices through the DM sheet in Eulerian space at $z=0$ after evolving from four different initial conditions, with $\sigma^2_2$, $\sigma^2_6$, $\sigma^2_{10}$ and $\sigma^2_{14}$, smoothed with the $\widetilde{W}_{\sk}$ filter. As $\sigma^2$ increases (more small scale $k$ modes included in the initial $P(k)$), more collapsed structures emerge on smaller scales - haloes (cyan), filaments (blue), pancakes (orange) and voids (white).}
    \label{fig:euler}
\end{figure}

\begin{figure}
    \includegraphics[width=\columnwidth]{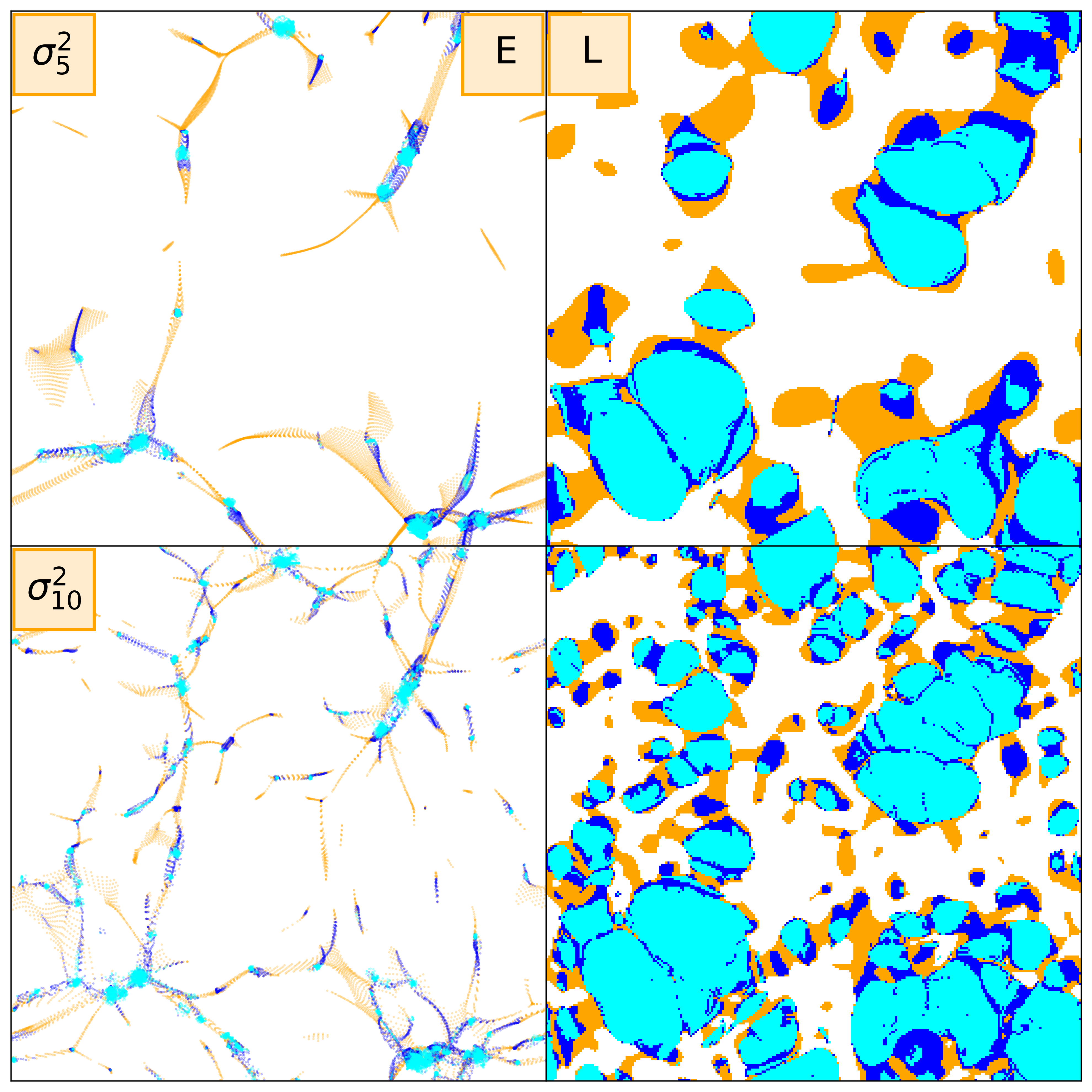}
    \caption{Two slices of $\widetilde{W}_{\sk}(kR)$ smoothed simulations, evolved from initial conditions with $\sigma^2_{5}$ (top) and $\sigma^2_{10}$ (middle). \textbf{Left:} the fluid element distribution in Eulerian space. \textbf{Right:} the same distribution, but in Lagrangian space. The colour scheme of the top four panels is the same as in Figure \ref{fig:euler}.}
    \label{fig:EL_sk}
\end{figure}

\begin{figure*}
    \includegraphics[width=0.9\textwidth]{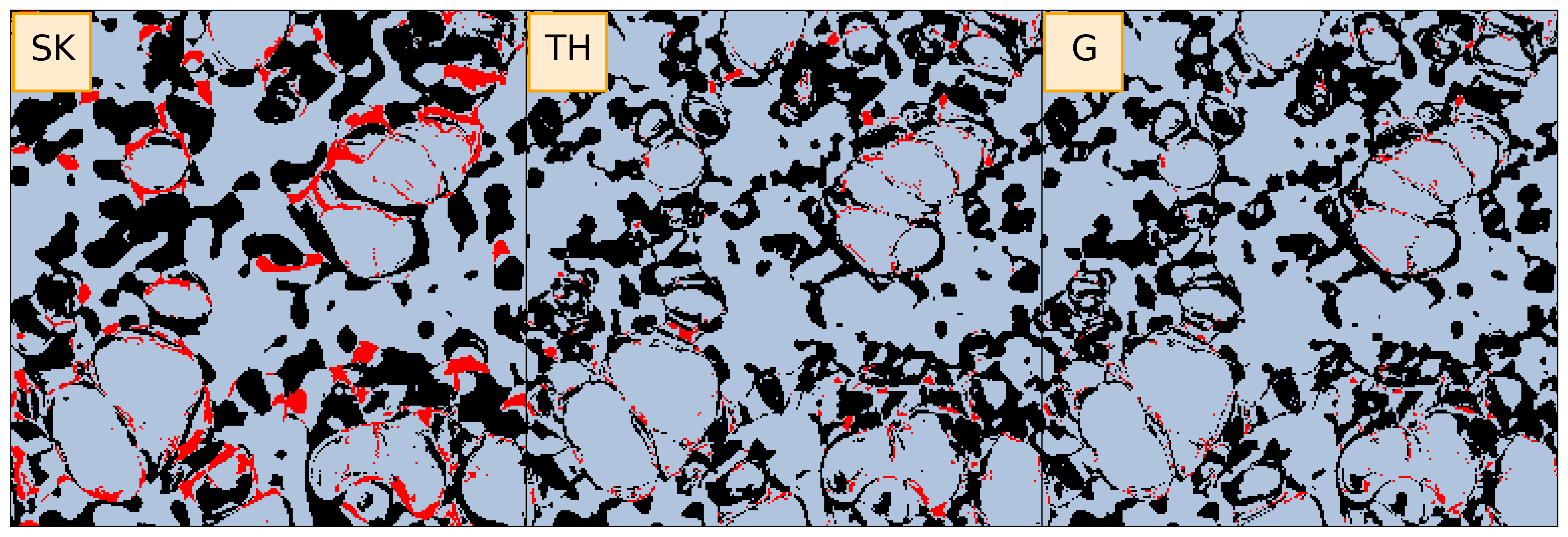}
    \caption{The difference of the two upper simulations $\sigma^2_{10} - \sigma^2_5$, but for all three window functions, $\widetilde{W}_{\sk}$, $\widetilde{W}_{\tth}$ and $\widetilde{W}_{\g}$, from left to right. The particle distribution is the same as in $\sigma^2_{10}$. Particles belonging to the same morphology class as in $\sigma^2_5$ are shaded grey, those which collapsed along more axes $n_{10} > n_5$ are marked in black and the cases where $n_{10} < n_5$, i.e. un-collapsed particles are coloured red.}
    \label{fig:uncollapse}
\end{figure*}

To illustrate this concept, consider Figure \ref{fig:scale-space-excursion}, where we plot the evolution of a Lagrangian field $\vec{q}$ with increasing smoothing scale $R$. In the left panel we present a well-defined excursion set example employing a perfect tri-axial collapse model, where $n_R$ consistently \textit{increases} with \textit{decreasing} $R$. The tri-axial nature is represented by the rank of each element, colour-coded to indicate its morphology type:  white for $n_R = 0$ \textit{(void)}, orange for $n_R = 1$ \textit{(pancake)}, blue for $n_R = 2$ \textit{(filament)}, and cyan for $n_R = 3$ \textit{(halo)}.  We will maintain this colour scheme to represent the morphology classes of fluid elements in all subsequent figures depicting our simulations. In the right panel, we demonstrate a contrasting example where the behavior in scale-space violates the excursion set assumption, as some regions of $n_R=1$ revert to $n_R=0$.  This panel also highlights instances of non-hierarchical collapse behavior, where elements transition directly from  $n_R = 0$ to $n_R = 2$ (see Section \ref{subsec:coll_criterion}) as $R$ decreases.

Further, in Figure \ref{fig:euler} we show the final snapshot ($z=0$) of the same slice through the evolved density field within our simulation box. Each slice has evolved from different initial conditions, with the sharp $k$-space cut-off $\widetilde{W}_{\sk}$ applied to the $P(k)$ (corresponding to $\sigma^2_2$, $\sigma^2_6$, $\sigma^2_{10}$ and $\sigma^2_{14}$ from the top-left to the bottom-right), as described in Section \ref{subsec:excsims}. These snapshots depict the final distributions of simulation particles at their current coordinates
$\vec{x}$ in the Eulerian frame. 

In general, decreasing the smoothing scale primarily introduces additional small-scale structure while preserving structure that has already existed on larger scales. Moreover, structures appear to move up in their morphology rank when the smoothing scale is decreased. For example, filaments may additionally fragment into smaller-scale haloes embedded in the larger filament. Pancakes, on the other hand, may fracture into smaller filaments and haloes. 

Figure \ref{fig:euler} suggests that the excursion set assumption holds well on a qualitative level. We will further evaluate this quantitatively in the following subsection.

\begin{figure*}
    \includegraphics[width=1\textwidth]{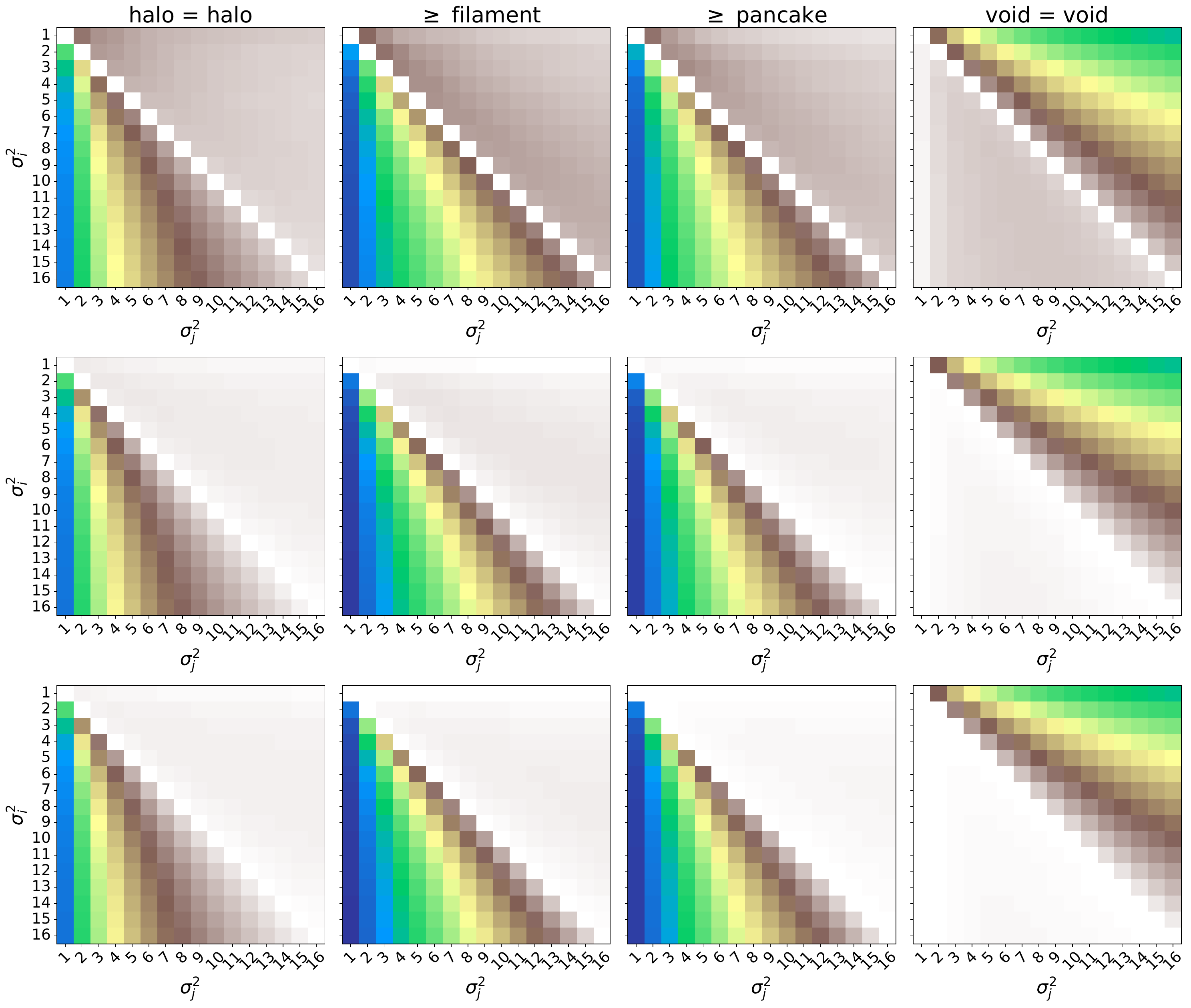}
	\includegraphics[width=0.8\textwidth]{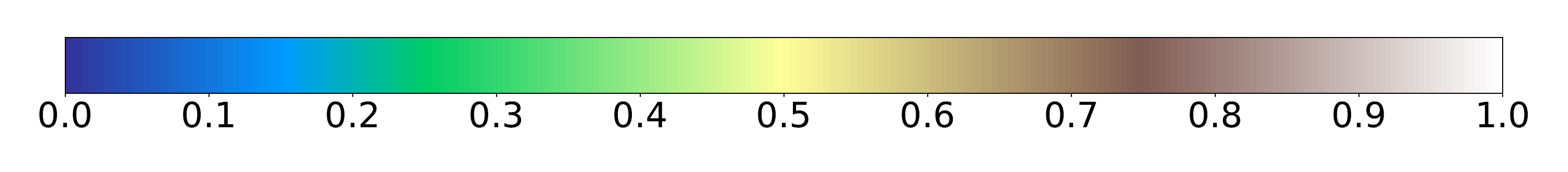}
    \caption{Four confusion matrices per filter: $\widetilde{W}_{\sk}$ (top), $\widetilde{W}_{\tth}$ (middle) and $\widetilde{W}_{\g}$ (bottom), each showing the total of particles from $\sigma^2_i$ run which are assigned to the same morphology, or to one of higher rank, in run $\sigma^2_j$. I.e. the "$\geq$ pancake" confusion matrix quantifies the fractions of pancake particles from $\sigma^2_i$ that are either pancakes, filaments or haloes in $\sigma^2_j$. Moving to the right of the main diagonal, the smoothing scale of the comparison simulation decreases, while moving to the left, the scale increases. The excursion set theory assumption implies that everything to the \textit{right} of the main diagonal, in the three plots from the left, should be white. Conversely, everything to the \textit{left} of the main diagonal in the "void = void" confusion matrix should also be white.  }
    \label{fig:conf_matrices}
\end{figure*}

\begin{figure}
	\includegraphics[width=\columnwidth]{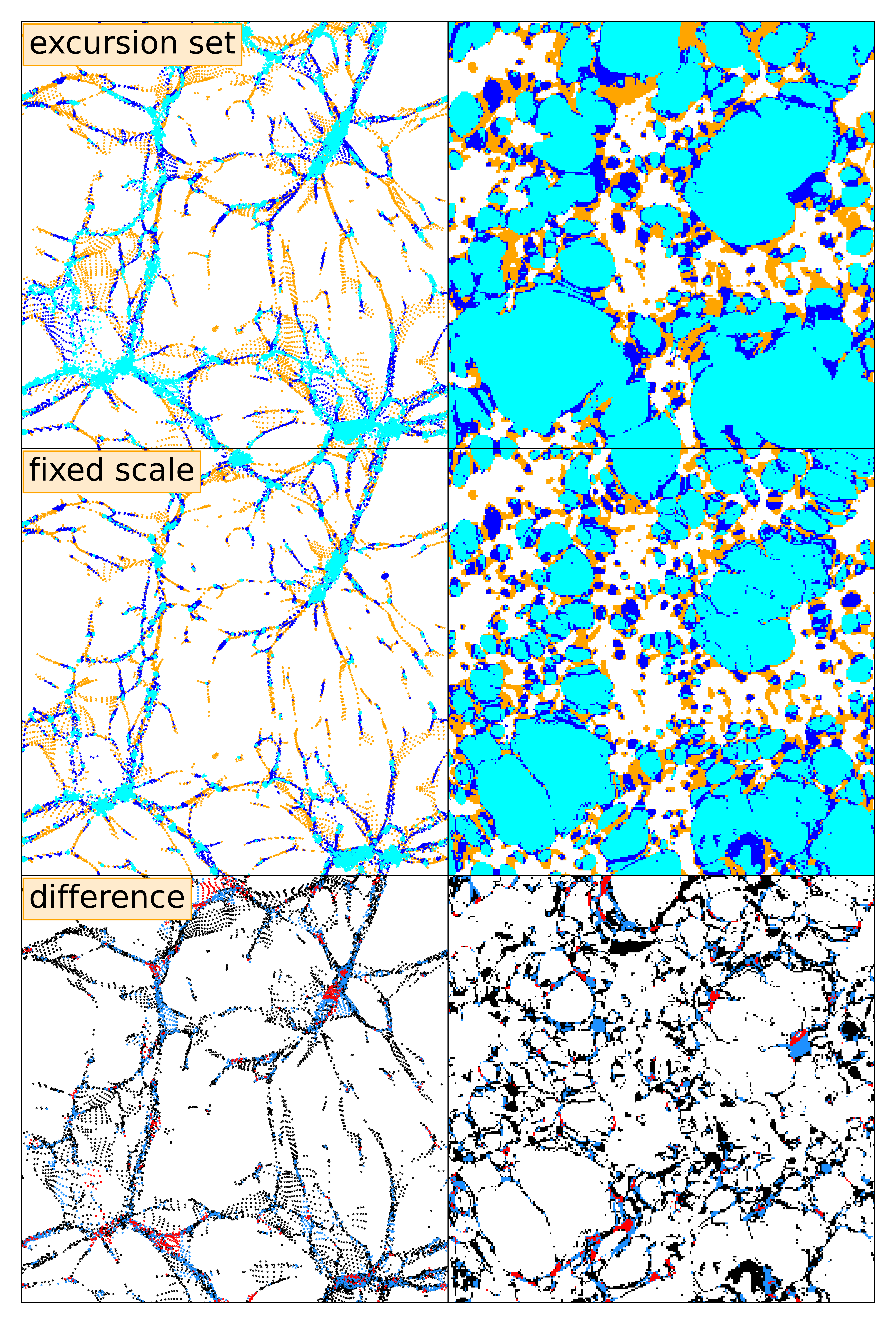}
    \caption{The excursion set constructed from the 16 simulations smoothed with $\widetilde{W}_{\sk}$ (top), the $\sigma^2_{16}$ output (middle) and their difference (bottom). The colour scheme of the difference plot is as follows:  white indicates no difference, while black, blue and red correspond to differences of $1,2 \, \text{and} \, 3$, respectively, in the morphology rank $n_R$. The simulation particles are plotted in Eulerian space (left panels) and Lagrangian space (right panels).}
    \label{fig:sim_ex_sk}
\end{figure}

\begin{figure}
	\includegraphics[width=\columnwidth]{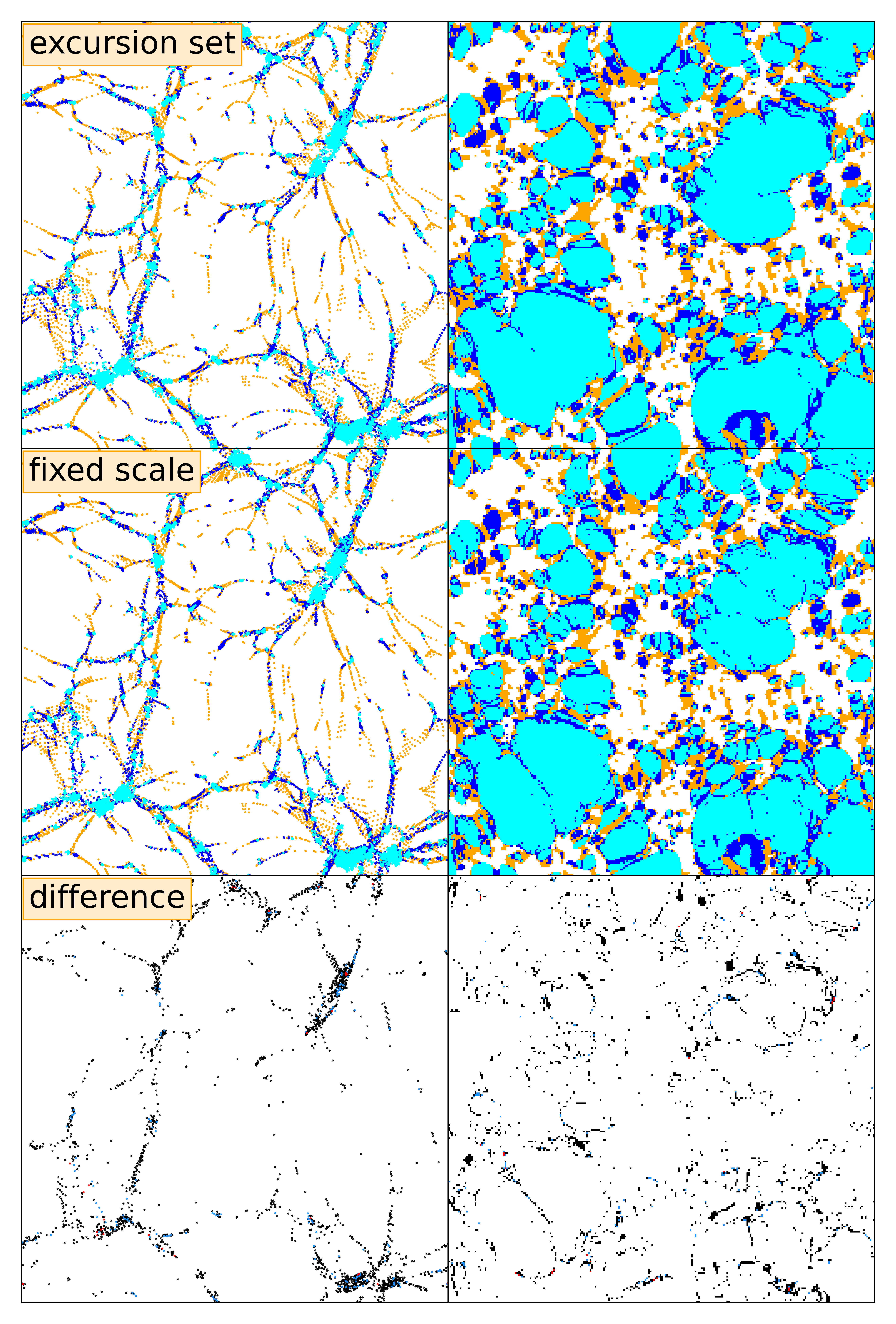}
    \caption{Same as Figure \ref{fig:sim_ex_sk}, but for simulation ICs smoothed with $\widetilde{W}_{\tth}$.}
    \label{fig:sim_ex_th}
\end{figure}

\begin{figure}
	\includegraphics[width=\columnwidth]{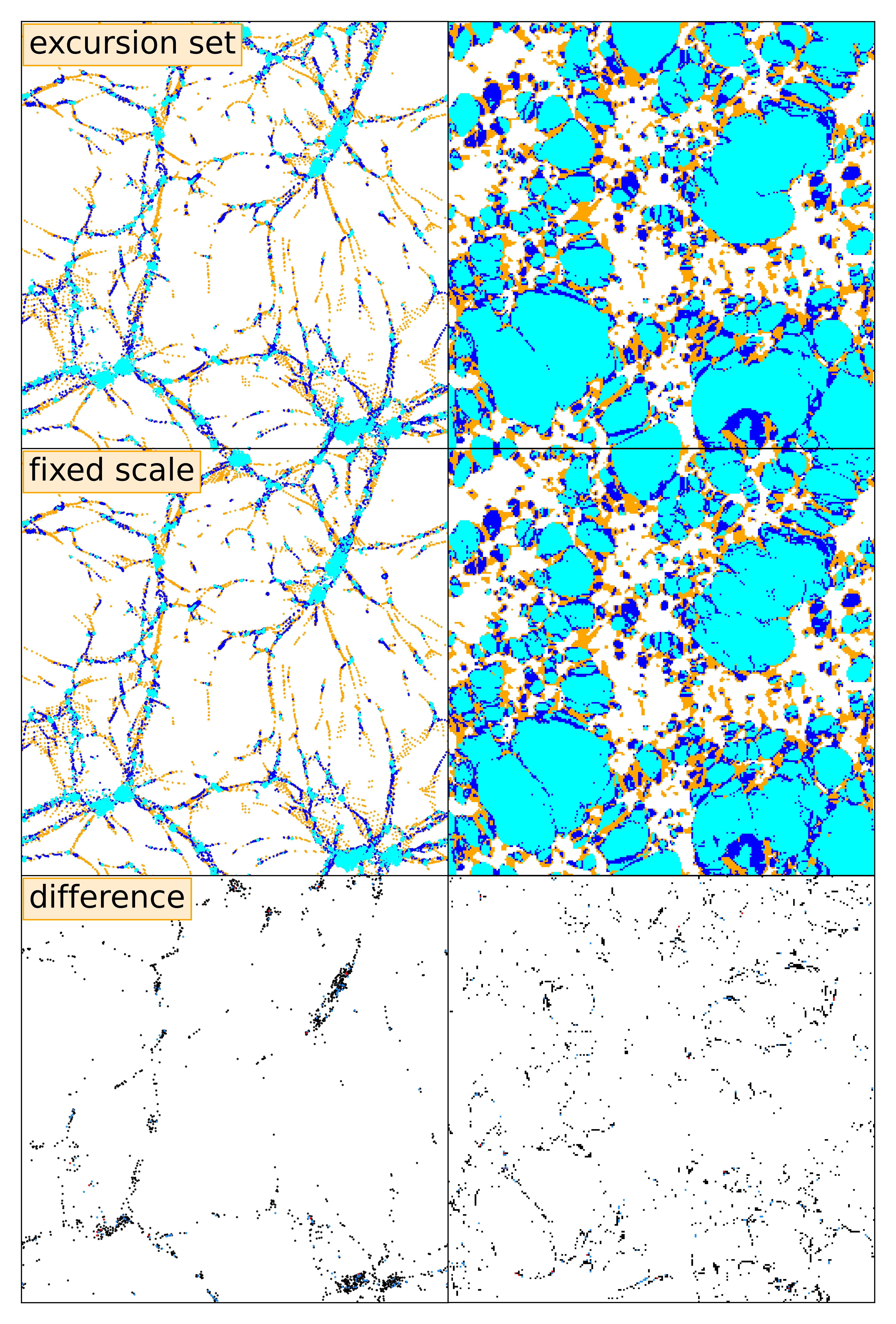}
    \caption{Same as Figure \ref{fig:sim_ex_sk}, but for simulation ICs smoothed with $\widetilde{W}_{\g}$.}
    \label{fig:sim_ex_gauss}
\end{figure}

\subsection{Testing the excursion set assumption}
\label{subsec:4.2}

Although Figure \ref{fig:euler} demonstrates in Eulerian space that structures typically tend to move upwards in their morphology rank when adding small-scale perturbations, we need to use a Lagrangian perspective to test Equation \eqref{eqn:excursion_inequality} quantitatively. 

For illustration, consider Figure \ref{fig:EL_sk}, where we present the morphology classification in the final snapshots of two different $\widetilde{W}_{\sk}$ smoothed simulations, $\sigma^2_5$ (top) and $\sigma^2_{10}$ (bottom). The panels on the left show the simulation points plotted in \textit{Eulerian} space (indicated with an "E"), while in the panels on the right, the points are plotted in \textit{Lagrangian} space (indicated with an "L") -- that is, at their locations in the initial conditions. The colours indicate morphology classification, as outlined in Section \ref{subsec:4.1}. Note that haloes take up the least volume in Eulerian space and voids the most. However, in Lagrangian space, haloes take up the most volume because they are the densest and most massive structures with the most particles. We again note that most structures that exist in Lagrangian space at $\sigma^2_5$ also exist in the $\sigma^2_{10}$ simulation.

Figure \ref{fig:uncollapse} shows the $\sigma^2_{10}$ distribution of particles in Lagrangian space only, but this time the colouring reflects whether that simulation exhibits additional collapse with respect to the $\sigma^2_{5}$ simulation: gray for \textit{no difference} ($n_{10} = n_{5}$), black for \textit{collapse} ($n_{10} > n_{5}$) and red for \textit{un-collapse} ($n_{10} < n_{5}$). Thus, the gray and black regions are consistent with the excursion set assumption, but the red regions are not expected. In the leftmost panel, corresponding to $\widetilde{W}_{\sk}$, we note that most of the volume is black or grey, showing that the assumption is overall quite reasonably satisfied. We observe similar effects in the $\widetilde{W}_{\tth}$ (middle) and $\widetilde{W}_{\g}$ (right) smoothed simulations, but with considerably smaller un-collapsed regions. However, we also note that a small fraction of Lagrangian space seems to un-collapse due to the perturbations from smaller scales -- a phenomenon that, by construction, cannot be captured in the excursion set paradigm. 

A closer examination of Figure \ref{fig:uncollapse} reveals that the un-collapsing regions are pre-dominantly located near the boundaries of haloes in the $\sigma^2_{5}$ simulation. That these regions are the most affected by smaller-scale perturbations is understandable since these particles are just at the threshold of reaching their halo by $z=0$, and a small difference in the initial conditions may push them over the edge. Further, the coherence scale of the difference field seems more closely related to the smoothing scale of the $\sigma^2_{5}$ than of the $\sigma^2_{10}$ simulation. We may, therefore, speculate that most of the un-collapsing happens due to the perturbations that lie just below the cut-off scale of the $\sigma^2_{5}$ simulation. This can be seen in more detail in Figure \ref{fig_app:uncoll}, where the same plot is shown for additional examples.

For a quantitative assessment, we analyse the complete set of confusion matrices corresponding to $\widetilde{W}_{\sk}$, $\widetilde{W}_{\tth}$ and $\widetilde{W}_{\g}$ from top to bottom in Figure \ref{fig:conf_matrices}. In each confusion matrix, each row corresponds to a distinct group of particles associated with a specific morphology in the $\sigma_i^2$ simulation (with $n_i = 3, 2, 1$ or $0$ for the respective column panels). Each pixel in the same row shows which fraction of these particles is also collapsed to a compatible structure in the $\sigma_j^2$ simulation. That means $n_j=3$ for haloes, $n_j \geq 2$ for filaments, $n_j \geq 1$ for pancakes and for voids we compare to $n_j = 0$.

Elements left of the diagonal ($\sigma_j^2 < \sigma_i^2$), show the progression of the excursion set: the more small-scale perturbations are added, the more likely it is that a point collapses. The closer $\sigma_j^2$ is to $\sigma_i^2$, the more likely it is that particles have collapsed to the same type of structure they were selected by at $\sigma_i^2$, until the fraction corresponds to 1 at the diagonal by definition. 

Elements right of the diagonal ($\sigma_j^2 > \sigma_i^2$), show the fraction of particles that have the same (or a consistent) morphology also at higher variance. For example, in the second column panel, we see particles that are in a filament at $\sigma_j^2 = 5$ and are also in a filament (or halo) at higher variances. According to the excursion set assumption, all of the elements to the right of the diagonal should be at $100\%$ for all filters. However, in the simulations smoothed with $\widetilde{W}_{\sk}$, we observe that they are approximately $90\%$ instead - indicating that around $10\%$ of the mass un-collapses from filaments to lower rank structures. This deviation from $100\%$ gradually improves as we look to the $\widetilde{W}_{\tth}$ and $\widetilde{W}_{\g}$ panels, where the un-collapsed fraction levels at $\sim 4\%$ and $\sim 2\%$ for the two filters respectively.

For the halo, filament, and pancake cases, the deviation of the elements at the top right of the diagonal from unity indicates the degree of the excursion set assumption violation. A key insight from from Figure \ref{fig:conf_matrices} regarding the $\widetilde{W}_{\sk}$ smoothed simulations is that the major un-collapsing events occur when $\Delta\sigma^2 = 1$, corresponding to the confusion matrix squares directly adjacent to the diagonal. 

Beyond this point the fractions remain relatively constant, indicating that perturbations on much smaller length scales indeed do not impact the collapse of larger scale regions. However, perturbations on slightly smaller length scales do affect their collapse. For the $\widetilde{W}_{\tth}$ and $\widetilde{W}_{\g}$ smoothed simulations, major un-collapsing events occur rather at $\Delta\sigma^2 = 2$, however with a maximum deviation from the diagonal of $5\%$, but mostly $4\%$ for the top-hat and $3\%$, but mostly $2\%$ for the Gaussian filters, with this fraction remaining largely stable across the smaller scale $\sigma^2_j$.

For voids, the diagram needs to be read in the reverse order. The excursion set assumption implies that everything that is part of a void at the scale $\sigma_i^2$, must also be void at all larger scales ($\sigma_j^2 < \sigma_i^2$). Therefore, all pixels to the left of the diagonal are expected to be at $100 \%$. As with the other column panels, we can see that the fractions are generally of order $90\%$ and that most of the difference comes from scales that are very close to the diagonal for the $\widetilde{W}_{{\sk}}$ smoothed simulations. In contrast, for $\widetilde{W}_{\tth}$ and $\widetilde{W}_{\g}$ cases, this difference is at most $2\%$ and $1\%$, respectively. 

\begin{center}
    \begin{table}
        \caption{The un-collapsed fractions of particles between the three excursion sets and the reference $\sigma^2_{16}$ of the three suites of smoothed simulations. $U$ indicates the number of un-collapsed particles of the given type X. The fractions are taken with respect to the number N of X particles collapsed in the excursion set and $N_\text{Total}=256^3$.}
        \label{tab:accuracy_table}
        \begin{tabular}{ |l|c|c|c| } % 5 columns, alignment for each
    	\hline
    	& $\widetilde{W}_\text{sk}$ & $\widetilde{W}_\text{th}$ & $\widetilde{W}_\text{g}$  \\
        \hline \hline
        X & \multicolumn{3}{|c|}{$U_\text{X}/N_\text{X}$} \\
        \hline
        Halo & 0.16 & 0.05 & 0.04 \\
        Filament & 0.48 & 0.1 & 0.06 \\
        Pancake & 0.44 & 0.06 & 0.04 \\
        \hline
        Total & 0.2 & 0.05 & 0.035 \\
        \hline
        \end{tabular}
    \end{table}
\end{center}

\subsection{The simulation excursion set}
\label{subsec: sim_excursion_set}

Since the simulation represents a perfect collapse model, we may ask what happens if we follow through with the excursion set procedure, but use the simulated morphology classification. Using the simulation classification as a collapse model in equation \eqref{eqn:excursionset},  we get the simulation excursion classification:
\begin{align}
    n_{R,\mathrm{ES}} &= \sup \{ n_{R'} \, | \, R' \geq R \}
\end{align}
Under the excursion set assumption, the relationship should hold as $n_{R,\mathrm{ES}} = n_R$.

Figures \ref{fig:sim_ex_sk} through \ref{fig:sim_ex_gauss} demonstrate the result of emulating this excursion set with simulations smoothed with all three windows in the two top panels of each figure, again with the Eulerian and Lagrangian slices left and right, respectively. The middle rows show the simulations with the smallest fixed scale smoothing $\sigma^2_{16}$ and the bottom panels show the difference between the excursion set and $\sigma^2_{16}$, with the following colour assignment: $n_{\text{ES}} - n_{16} = 0$ (white), $n_{\text{ES}} - n_{16} = 1, 2, 3$ (black, blue, red), with ES denoting excursion set. As anticipated from the previous results, the observation of un-collapsing throughout the 16 runs indicates that the excursion set of the $\widetilde{W}_{\sk}$ smoothed simulations contains many more collapsed structures. While voids largely remain intact, haloes, filaments and pancakes are notably more spread out and smoother in the excursion set plot shown in Figure \ref{fig:sim_ex_sk}. Overall, the excursion set generated from these simulations results in a $20 \%$ of the total un-collapsed particle fraction relative to the $\sigma^2_{16}$ realisation. In this example, we also observe that particles identified as haloes in the excursion set are the least affected by changes to the ICs, with $16 \%$ undergoing un-collapse with respect to $\sigma^2_{16}$. In contrast, as many as $44 \%$ of pancakes and $48 \%$ of filaments in the excursion set experience un-collapse compared to $\sigma^2_{16}$. Similarly, in line with the findings from Figure \ref{fig:conf_matrices}, we observe a decrease of the overall un-collapsed regions when comparing the excursion sets generated from the top-hat (Figure \ref{fig:sim_ex_th}) and the Gaussian (Figure \ref{fig:sim_ex_gauss}) smoothed simulations with their respective $\sigma^2_{16}$. While the total fraction of particles undergoing un-collapse between the top-hat excursion set and the fixed scale simulation is $5\%$, this value is even lower in the Gaussian window example, amounting to just $3.5\%$. When focusing only on the fraction of particles classified as haloes, filaments or pancakes in the excursion sets that un-collapse in $\sigma^2_{16}$, we observe differences of $5\%$, $10\%$ and $6\%$ for the $\widetilde{W}_{\tth}$ case and $4\%$, $6\%$ and $4\%$ for the $\widetilde{W}_{\g}$ case, respectively. As in the $\widetilde{W}_{\sk}$ example, haloes are the least affected by changinges in the ICs, while filaments are the most impacted. In both instances, the smoothing of the collapsed structures in the excursion sets is hardly visible in the figures and is predominantly observed around haloes, while in the bottom panel of Figure \ref{fig:sim_ex_sk}, it is evident that the filament and pancake regions are also significantly affected by un-collapse. These values are summarised in Table \ref{tab:accuracy_table}. 

We conclude that the excursion set assumption does draw an appropriate picture on a qualitative level and that it also works reasonably well with the $\widetilde{W}_{\sk}$ smoothing and surprisingly well with $\widetilde{W}_{\tth}$ and $\widetilde{W}_{\g}$ smoothing on a quantitative level. In this section, we demonstrated that the core assumptions of the excursion set hold extremely well in practice with a Gaussian window, slightly worse with a top-hat and significantly worse with a sharp $k-$space window functions. That is, if we combine a perfect collapse model (a simulation) with a sharp $k$-space filtering method, then we find quantitative deviations from the core assumption on the particle-by-particle basis of order $20\%$. This implies that an excursion set formalism using $\widetilde{W}_{\sk}$ for smoothing, along with a maximally physical collapse model, \textit{should} yield a $20\%$ error.

\begin{figure}
	\includegraphics[width=1\columnwidth]{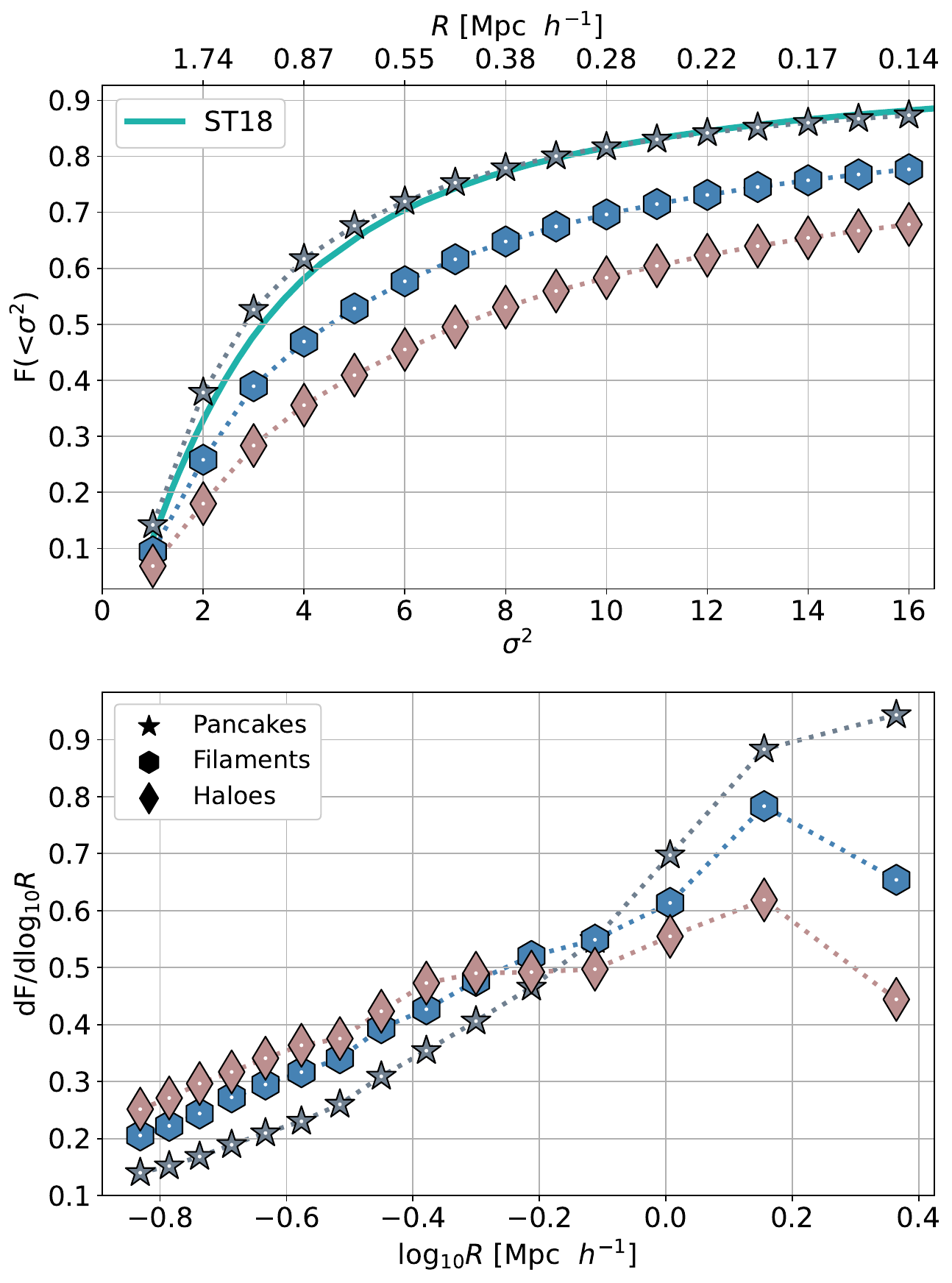}
    \caption{\textbf{Top:} The mass fractions of particles in pancakes (stars), filaments (hexagons) and haloes (diamonds) as a function of the smoothing scale $\sigma^2$ and $R$. The turquoise curve represents $F(<\sigma^2)$ of collapsed particles from \citet{stucker2018median}. \textbf{Bottom:} the particle fraction in structures of the ES radii. Both panels show the distributions for the $\widetilde{W}_{\sk}$ smoothed excursion set of simulations.}
    \label{fig:Mass_fractions}
\end{figure}

\begin{figure*}
	\includegraphics[width=1\textwidth]{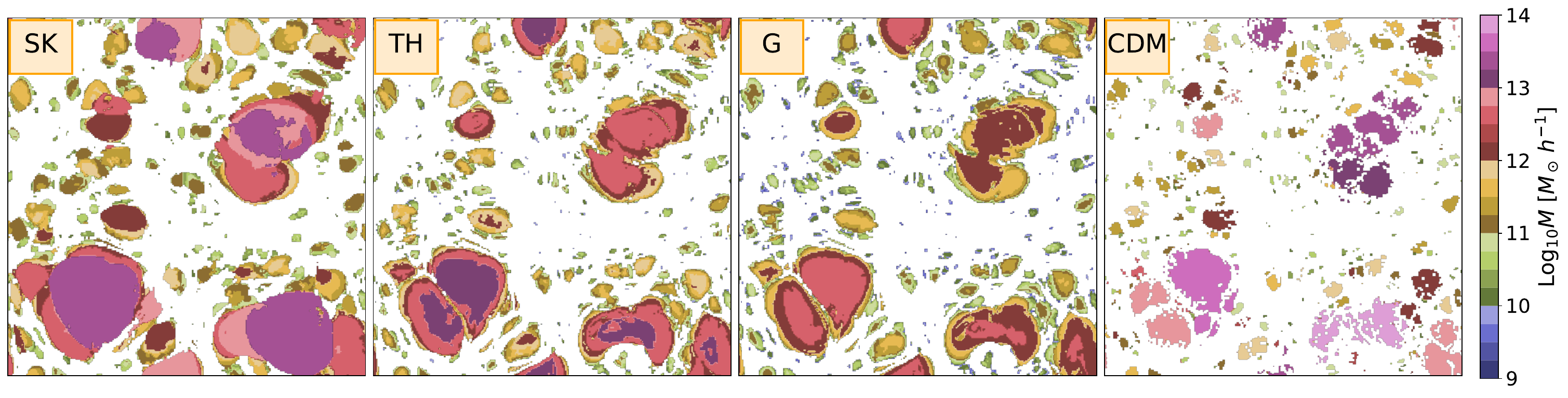}
    \caption{Lagrangian plots of excursion set mass assignment to simulation particles collapsed into haloes, based on the scale of collapse $\sigma^2_i$ and by the excursion set assumption. The first three panels show the results from the suite of simulations ran with initial $P(k)$ smoothed with the $\widetilde{W}_{\sk}$, $\widetilde{W}_{\tth}$ and $\widetilde{W}_{\g}$, from left to right. The rightmost panel shows the {\tt FoF} haloes from a CDM (no smoothing) simulation.} 
    \label{fig:Mass_excursion}
\end{figure*}

\subsection{Morphology mass fractions}
\label{subsec: mass_functions}
So far, we have focused on the particle-by-particle contrast of the excursion sets generated with different window functions and the fixed scale $\sigma^2_{16}$ simulations. As already discussed in the Introduction, the excursion set formalism has many other useful applications, such as predicting HMFs, which offer valuable insights into the cosmic mass distribution within haloes at all scales relevant to the chosen cosmology. This methodology can also be extended to explore the mass distribution in other structures, such as pancakes and filaments. However, gaining a more comprehensive understanding requires data that is not easily accessible through approaches relying on standard structure finders. Using our simulations, we can not only identify the morphologies of the structures into which the simulation particles collapse, but also obtain a general understanding of the scales at which these structures form when combined with the excursion set framework.

We calculate the cumulative mass fraction $F(<\sigma^2)$ or $F(>R)$ for each morphology class. We do this by identifying the scale of the first (and largest) collapse in the excursion set, as defined by eqation (\ref{eq:sim_ES_def}) and determining the fraction of particles with a collapse scale larger than the considered scale. Recount that in our way of counting here, every particle embedded in a filament will also be considered part of a pancake, and so on.

The top panel of Figure \ref{fig:Mass_fractions} shows the cumulative distributions of particle fractions associated with pancakes (stars), filaments (hexagons) and haloes (diamonds) based on the excursion set of simulations obtained with the $\widetilde{W}_{\sk}$ smoothing \footnote{We have checked that other kernels lead to similar results, but for simplicity we focus only on the sharp$-k$ case here.}. In the plot, the upper x-axis indicates the Lagrangian length smoothing scales, while the lower x-axis  shows the corresponding values of $\sigma^2$. We observe that typical particles appear to be part of a large-scale pancake, a smaller-scale filament and a notably smaller halo. For example the median particle (where $F=0.5$) is part of a $R \sim 1.8$ Mpc $h^{-1}$ pancake, a $R \sim 0.8$ Mpc $h^{-1}$ filament and a $R \sim 0.5$ Mpc $h^{-1}$ halo. In Appendix \ref{app: mass_fractions} we also show the same plot, bur for the example of the $\widetilde{W}_{\tth}$ smoothed simulations.

At the smallest considered smoothing scale ($R \sim 0.14 \, \mathrm{Mpc}\,h^{-1}$) -- which is already considerably smaller than what most cosmological simulations resolve -- the morphology fractions for pancakes, filaments and haloes are $88\%$, $78\%$ and $68\%$ respectively. Since the mass fractions are growing moderately in this regime, we would expect that in most cosmological simulations of order $80-90\%$ of particles are embedded in a pancake (or higher), $70-80\%$ in a filament (or halo) and $60-70\%$ in a halo. The exact numbers may naturally vary based on the specifics of the classification \citep[see][for a comprehensive review]{2011haloFinders, 2013Knebe, AnguloHahn2022}. Additionally, we anticipate these values to rise considerably if simulations were extended to the physical cut-off in the power spectrum, such as in a neutralino cosmology \citep{stucker2018median, wang_2020_vvv}.

We can compare our results to existing predictions from excursion sets. In particular, the mass fractions of pancakes and filaments were previously explored by \citet{shen2006excursion}. In this work, excursion sets in density with three separate barriers for pancakes, filaments and haloes (based on approximations of the ellipsoidal collapse model) were used to explore the formation of such structures. They find in a $\Lambda$CDM Universe, at $z=0$, above $99\%$ of all mass is contained in pancakes, $72\%$ in filaments and $46\%$ in haloes, with $M > 10^{10} \, \M_{\odot}$. Since their resolved mass is exactly in the range of the sharp$-k$ mass at our smallest resolved scale $\sim 2 \times 10^{10} \, \M_{\odot}$, we can compare these numbers approximately to ours. Clearly the model of \citet{shen2006excursion} strongly over-predicts the mass in pancakes. This is also confirmed by a variety of other studies based on cosmic web classification schemes \citep{veldisp2019}, which tend to find numbers between $11\%$ and $32.5\%$ for the mass in structures with $n \geq 1$.

In the top panel of Figure \ref{fig:Mass_fractions}, we also plot $F(<\sigma^2)$ adapted from Figure 9 of \citet{stucker2018median} (turquoise). Their triaxial collapse model has been designed to describe the first collapse of voids towards pancakes (or single-stream regions to multi-stream regions). The authors used random walks of the deformation tensor to determine statistics of single-stream regions ($n=0$).  The fraction of mass in pancakes -- given by the remaining collapsed material -- appears to be very well predicted by this approach. Consequently, it can be concluded that, at the very least, an accurate description of the full hierarchy of three-dimensional collapse requires accounting for the full deformation tensor and is challenging to achieve using effective density barriers, as is done for haloes \citep{sheth2001}. 

The bottom panel of Figure \ref{fig:Mass_fractions} shows the differential mass fractions, represented by the first crossing distribution $\mathrm{d}F / \mathrm{d} \log_{10} R$, as a function of scale. It reveals that large-scale pancakes appear more significant compared to haloes, whereas small-scale pancakes appear less dominant. We may conclude that most of the smaller-scale web patterns of the universe will likely be accreted into larger-scale pancakes.

\subsection{Collapse scale to mass relation} 
\label{subsec: ES_mass_ass}
So far, we have tested the general assumptions of excursion set formalisms, particularly in terms of how the hierarchy of structure formation progresses across scale-space. However, excursion sets are very commonly used to predict the mass functions of haloes, which requires an additional assumption. It is typically assumed that when a particle first satisfies a collapse criterion at a smoothing scale $R_s$, it becomes part of a halo with mass $M \propto R_s^3$, as outlined in equation \eqref{eq:mass}. This mapping between collapse scales and the assumed halo mass is a crucial ingredient that we can examine with our excursion set of simulations.

Consider Figure \ref{fig:Mass_excursion}, where we label each particle in Lagrangian space by the predicted halo mass, based on the default mass mapping \eqref{eq:mass}, for the excursion sets obtained from the simulations that have been smoothed with the different filter functions. Additionally, in the rightmost panel, we also show each particle labeled by the mass $\M_{\mathrm{FoF}}$ of the halo that it has been found for a CDM (no smoothing) simulation with a friends-of-friends halo finder with linking length 0.16.

We note that our collapse criterion is more permissive than the traditional friends-of-friends definition, which typically includes all particles up to the splash-back radius. Further, we point out that the excursion sets may label particles that are part of the same structure with different masses, because some particles have only become part of the same structure at smaller smoothing scales. It is clear that the true collapse scale-to-mass relation is not a simple one-to-one relation, and is likely to be significantly more complex. In the Figure, the excursion set masses correlate reasonably well with the actual halo masses, but deviations are quite large in amplitude. 

\begin{figure*}
	\includegraphics[width=1\textwidth]{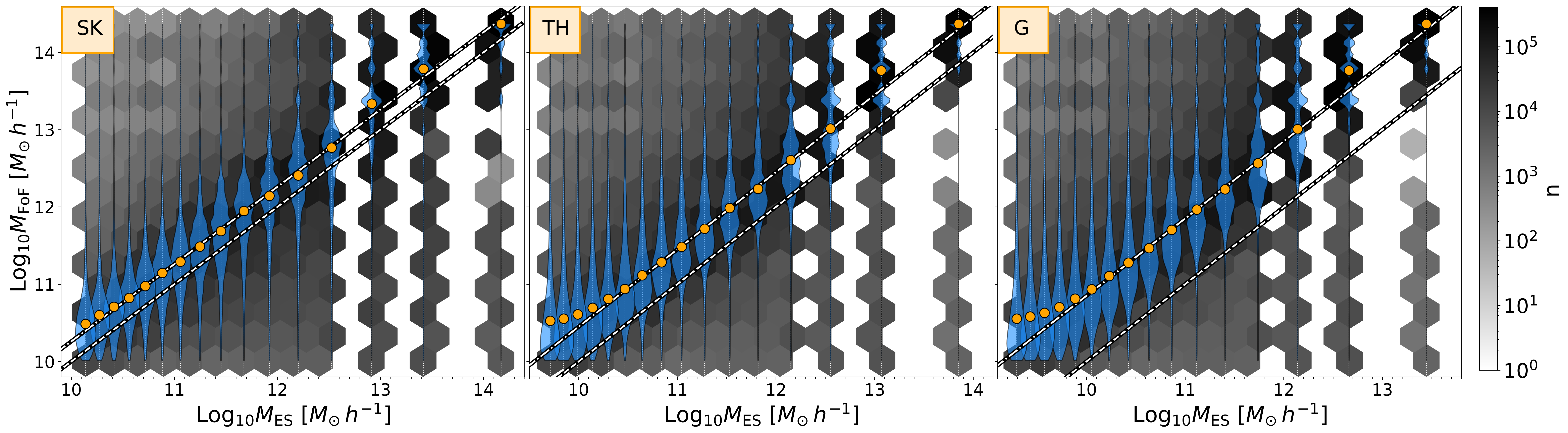}
    \caption{ The background hexagonal plot shows the particle count $n$ in the 2D bins, while the violin plots (blue) show the count distribution along the y-axis in the 16 bins corresponding to each $M_{\text{ES}}$. Medians of the $M_{\text{FoF}}$ distributions wrt to the $M_{\text{ES}}$ (orange points). The benchmark masses (white dashed) and the diagonal shifted along the y-axis by a factor $\alpha$ to match the medians (white dash-dotted).}
    \label{fig:violin}
\end{figure*}

To evaluate this more quantitatively, in Figure \ref{fig:violin} we show the relation between the excursion set collapse scale (indicated through mass, as in equation (\ref{eq:mass})) versus the measured masses of haloes of the same particles in a CDM simulation. We have to restrict this comparison to particles that are both part of a friends-of-friends halo in the simulation and have collapsed according to the simulation excursion set. We show this in two ways: (1) the greyscale, background hex bin distribution of particle counts, with finer 2-dimensional binning and (2) blue violin distributions of 6 coarsely binned counts along the x-axis. The orange circles indicate the medians of particle counts in the x-axis bins, and the diagonal, white-dashed lines show the benchmark M$_{\mathrm{FoF}}$ masses. Notice that the excursion set mass distribution is not continuous (we have exactly 16 masses corresponding to our sets of smoothed simulations), and due to our choice of y-axis binning, there are horizontal gaps in the hexbin distribution. Likewise, due to our small simulation volume ($L = 40$ Mpc), there is a lack of particle counts in high mass haloes ($ > 10^{13} \, \text{M}_{\odot} h^{-1}$) in the simulations, giving rise to horizontal data gaps. 

\begin{figure}
	\includegraphics[width=1\columnwidth]{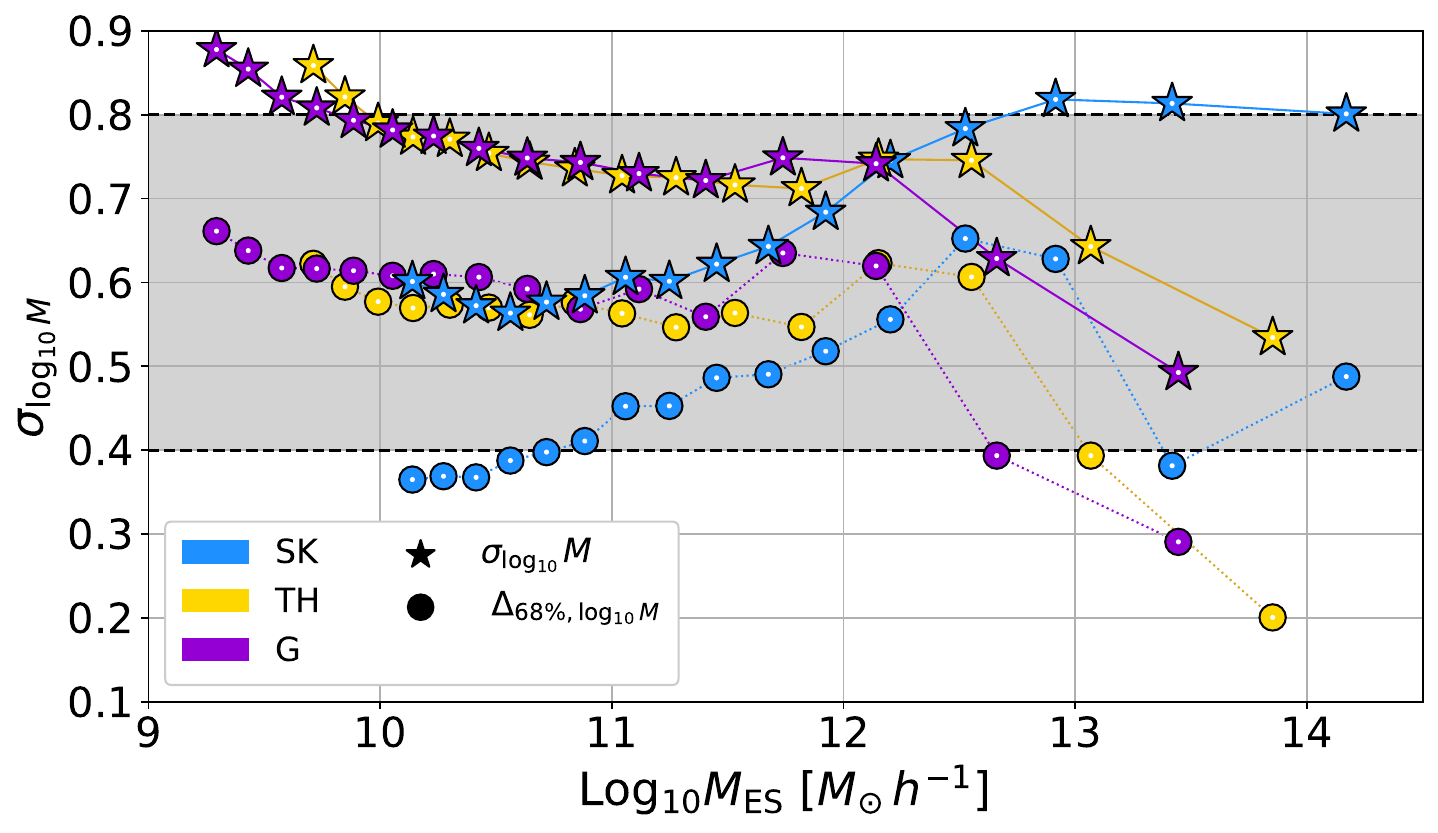}
    \caption{Two estimators of the logarithmic scatter in the mass-map. Stars indicate the standard deviation, and circles are a percentile-based estimate. While the precise amplitude depends on the details of the filter and the estimator, the overall amplitude ranges between $0.4$ and $0.8$ dex (shaded area).}
    \label{fig:Mass_scatter}
\end{figure}

We find two key insights from this figure: (1) The median relation tends to have an offset from the naive expectation from equation (\ref{eq:mass}). For the considered window functions the offset is $\alpha= 0.26, 0.45, 0.85$ for $\widetilde{W}_{\sk}$, $\widetilde{W}_{\tth}$ and $\widetilde{W}_{\g}$ respectively. We suggest that the prefactor of the mass map should generally be considered as a degree of freedom that may be fitted in excursion set models. This may account for the uncertainty associated with the halo definition and the effective scale of the smoothing function.
(2) The collapse scale-to-mass relation is much more complicated than commonly assumed. Particles that collapse at a singular smoothing scale often end up in haloes of different masses \citep[see also][]{sheth2001, 2002MonacoPinocchio2, 2013Achitouv, HahnParanjape:2014}. This will make it challenging to reliably predict masses, even when adopting a ``perfect'' collapse model. 

We evaluate this scatter quantitatively in Figure \ref{fig:Mass_scatter}. Here, we show two estimates of the scatter amplitude of $M_{\mathrm{FoF}}$ in fixed bins of $M_{\mathrm{ES}}$. The first estimate is the standard deviation in log-space 
\begin{align}
    \sigma_{\log_{\mathrm{10} } M} = \sqrt{\langle (\log_{\mathrm{10}} M - \langle \log_{\mathrm{10}} M \rangle)^2 \rangle}
\end{align}
and the second is half the extent of the $68\%$ interval 
\begin{align}
    \Delta_{68\%, \log_{\mathrm{10}}M} = \frac{1}{2}(\log_{\mathrm{10}} M_{84\%} - \log_{\mathrm{10}} M_{16\%})
\end{align}
where $M_{84\%}$ and $M_{16\%}$ are the $84$th and $16$th percentiles respectively. If $M_{\mathrm{FoF}}$ would follow a log-normal distribution at fixed $M_{\mathrm{ES}}$, then the two estimators should lead to identical results. 

We find that the estimator $\Delta_{68\%, \log_{\mathrm{10}}M}$ generally gives smaller values than $\sigma_{\log_{\mathrm{10} } M}$ by about $0.2$ dex, showing that the error distribution is not perfectly log-normal. Overall, amplitudes range between $0.4$ and $0.8$ dex, and the scatter seems to be relatively constant with mass, generally not changing by more than $0.2$ dex in the range below $M \lesssim 10^{13} \M_\odot h^{-1}$. The scatter can be even a bit smaller at higher masses, but the measurement is likely unreliable here due to finite-size effects and a very limited sample. Interestingly, the $\widetilde{W}_\sk$ leads to a slightly smaller scatter in the mass map than the other filters at masses $M \sim 10^{11} \M_\odot h^{-1}$.

Independently of these details, we may summarize that excursion sets with a ``perfect'' collapse model still have at least an uncertainty of order $0.4 - 0.8$ dex in the mass map. \citet{lucie_smith_2024} reported a similar mass scatter within this range (for $10^{11} - 10^{13.4} \, \text{M}_\odot \, h^{-1}$ haloes), in their halo collapse model based on convolutional neural networks \citep[see also][]{2024Dani}. Improvements in the collapse model cannot reduce this uncertainty; instead, it is rather due to the limitations imposed by the spherical kernel that is used to traverse scale-space \citep[see also][]{HahnParanjape:2014,lucie_smith_2024}.

\section{Discussion and Conclusion}
\label{sec:discussion_conclusion}
Excursion sets are an invaluable tool to predict and understand the formation of structures. While they are primarily utilized to predict statistical quantities, such as HMFs, they may also be used to infer detailed particle-by-particle predictions, such as those related to accretion and merger histories, as well as the formation of all types of structures like voids, pancakes, filaments and haloes. Such predictions are crucial for understanding structure formation beyond scales that can be resolved reliably in simulations. However, to make such predictions, excursion sets must be tested thoroughly at scales that \textit{can} be resolved in simulations.

Historically, excursion sets have been primarily evaluated based on their ability to reproduce the simulated HMF. However, in the era of precision cosmology, they have taken a backseat role to simulation-based fits \citep[e.g.][]{sheth2001, 2001Jenkins, 2003Reed, Reed_2006, 2006Waren, Tinker2008, 2016Diespali} and HMF emulators \citep[e.g.][]{2002MonacoPinocchio1, 2002MonacoPinocchio2, 2020AMiraTitan}. These simulation-derived HMFs typically achieve an accuracy better than $10 \%$ for all common halo mass definitions and can include possible non-universal behaviour of the HMF. It must be stressed that, by definition, these mass functions are limited only to the regime where simulations can be preformed and, as such, have little to no predictive capability beyond their calibrated range, consequently yielding no analytical insights. Predictions of unresolved quantities -- such as the abundance of very low-mass haloes -- thus still rests on excursion set models. If their accuracy were improved, these models could yield very economic alternatives to the large simulations required to calibrate high-accuracy mass functions. Such development would facilitate quick exploration of alternative cosmological models. 

The aspect of modeling, that has typically been optimized and improved upon, is the model describing the collapse of Lagrangian volume elements. Examples include spherical collapse \citep{press_1974, white_1979, bond1991excursion}, 3D ellipsoidal collapse \citep{bond_myers_1996, sandvik2007}, effective density-based ellipsoidal collapse \citep{sheth2001, shen2006excursion}, triaxial collapse \citep{1995Monaco, 2002MonacoPinocchio1, 2002MonacoPinocchio2, stucker2018median}, energy-based spherical collapse \citep{musso_2021} and virial-equation based collapse \citep{musso_2024}.

In this paper, we have shown that excursion sets can also be created using a ``perfect'' collapse model, namely, a simulation. This requires running several simulations with initial conditions smoothed on successively reduced scales. Collapse may be detected using conventional methods like a friends-of-friends classification or more advanced methods that distinguish between different structure morphologies, as we have adopted here.

This ``excursion set of simulations'' has allowed us to: (1) test assumptions of the excursion set models independently of the adopted collapse model, (2) define an upper limit to the possible realism of excursion set models, and (3) identify bottlenecks in the current paradigm. 

In this paper, we have tested two core assumptions of excursion sets. The first assumption states that \textit{collapse does not revert when decreasing Lagrangian smoothing scales}. We find that this assumption is overall satisfied surprisingly well and the quantitative accuracy depends significantly on the applied window function. For the sharp $k-$space filter, we find that up to 20\% of fluid elements un-collapse across the resolved scales, whereas for the Gaussian and top-hat windows, this percentage is only of order 3.5\% and 5\%. This degree of accuracy may be a reassurance for any models attempting to disentangle structure formation in scale-space. However, we also note that the best possible excursion set formalism (or Peak model) may not surpass this level of accuracy unless a more optimal window function or other degrees of freedom are chosen to compensate for this effect. For instance, it would be interesting to carry out a particle-by-particle analysis with the smooth $k-$space filter proposed by \citet{2018Leo-smoothKfilter}. They smooth the sharp transition from $kR$ to $0$ of the $\widetilde{W}_\sk$ by introducing a free parameter $\beta$. Notably, the HMF prediction with this filter only improves that of the sharp $k-$space filter for WDM with free-streaming mass in the order of $10^{10} \, M_\odot \, h^{-1}$, corresponding to $\sim 1$ keV DM and the smallest smoothing mass we calculate for the $\widetilde{W}_\sk$\footnote{They use the top-hat mass-mapping relation with an added free parameter $c$ (also calibrated with simulations), yielding slightly different $M$ values.} (Figure \ref{fig:P_cuts}). While their analysis focuses only on improving the HMF predictions, we anticipate it might, at best, increase particle-by-particle accuracy to a level comparable to that of the top-hat and Gaussian filters.

The second assumption we tested concerns the \textit{simple deterministic mass-mapping relation} (Equation \eqref{eq:mass}), that accounts for the amount of mass enclosed within a (spherically symmetric) filter scale. This process requires mapping the Lagrangian smoothing scale -- where the collapse criterion is first fulfilled -- to an estimated halo mass and is essential for inferring HMFs. By comparing the collapse smoothing scales with actual halo masses from a CDM simulation, we found two key insights: 
\begin{enumerate}
    \item \textit{Deviations from the standard mass relation}: The proportionality constant in the conventional $M \propto R^3$ relation often deviates from naive expectations and should be treated as a degree of freedom in excursion set formalisms. For example, reports of excursion set models performing better with collapse barriers below the standard spherical collapse threshold $\delta_c = 1.686$ \citep[e.g.][]{Delos2024} may stem from misattribution this degree of freedom. Alternative parameterisations of this relation have also been proposed -- such as those for sharp ad smooth $k-$space filters \citep[e.g.][]{2015SchneiderWDM, 2018Leo-smoothKfilter} -- but still require calibration against simulations. Additionally, \citet{HahnParanjape:2014} found a strong correlation between the scatter in excursion set-predicted masses (at fixed halo masses) and the shape of Lagrangian patches. Particularly, they found them to be mostly aspherical for low-mass, and spherical for high-mass haloes, which raises questions about focusing on spherically symmetric smoothing functions in excursion set analysis \citep[see also][]{2014Achitouv}.
    
    \item \textit{Significant uncertainty in the mass relation}: The mass-mapping relation exhibits substantial uncertainty of order 0.4 to 0.8 dex. This stochasticity effectively introduces additional smoothing in the mass-weighted HMF. A similar observation was made by \citet{HahnParanjape:2014}, who incorporated mass scatter into a modified HMF expression, achieving better agreement with simulations.  In future work we could address these insights by running simulations in a larger volume to increase the sample of high-mass halos ($\geq 10^{13} \, M_\odot \, h^{-1}$), confirming the extent of mass scatter at these scales.  While this effect is unlikely to significantly impact HMF predictions in CDM universes, it could be crucial in cases where the mass function has sharp features, such as those in universes with a power spectrum cut-off. Whether the excursion set theory can be improved in such cosmologies by including this effect will be interesting to investigate in future studies. 
\end{enumerate}
These insights highlight the necessity for further research in order to understand how the halo mass of a collapsed fluid element can be efficiently determined from additional simple criteria.

Lastly, we have also used our excursion sets to measure the Lagrangian scales of pancakes, filaments and haloes that typical fluid elements may be part of. Interestingly, the typical Lagrangian scales of pancakes tend to be significantly larger than those of haloes and filaments. Approximately $50\%$ of particles will be part of a halo with $R \gtrsim 0.5$ Mpc $h^{-1}$, a filament with $R \gtrsim 0.8$ Mpc $h^{-1}$ and a pancake with $R \gtrsim 1.8$ Mpc $h^{-1}$. Our measurements can serve as a benchmark for anisotropic collapse models \footnote{The values quoted here are for the sharp$-k$ window, however, other kernels lead to similar results.}, and they suggest that an appropriate treatment of the cosmic web requires collapse models based on the full deformation tensor, rather than effective density-based descriptions \citep{shen2006excursion}. 

Based on the findings of our investigation, we conclude that testing the detailed predictions of excursion set models against simulations may guide enhanced models. We will continue to explore these avenues in future work, where we will investigate the extent to which the approximate collapse models can accurately represent the triaxial evolution of a cosmic fluid element.

\section*{Acknowledgements}
JS acknowledges support from the Austrian Science Fund (FWF) under the ESPRIT project number ESP 705-N. REA acknowledges support from project PID2021-128338NB-I00 from the Spanish Ministry of Science and support from the European Research Executive Agency HORIZON-MSCA-2021-SE-01 Research and Innovation program under the Marie Skłodowska-Curie grant agreement number 101086388 (LACEGAL). The computational results presented have been achieved using the Vienna Scientific Cluster (VSC).

%%%%%%%%%%%%%%%%%%%%%%%%%%%%%%%%%%%%%%%%%%%%%%%%%%
\section*{Data Availability}
All data used in this article are available upon reasonable request from the corresponding author.



\bibliographystyle{mnras}
\bibliography{citations} % if your bibtex file is called example.bib

%%%%%%%%%%%%%%%%%%%%%%%%%%%%%%%%%%%%%%%%%%%%%%%%%%

%%%%%%%%%%%%%%%%% APPENDICES %%%%%%%%%%%%%%%%%%%%%

\appendix

\section{Un-collapse}
\label{app: uncollapse}

\begin{figure*}
    \includegraphics[width=0.6\textwidth]{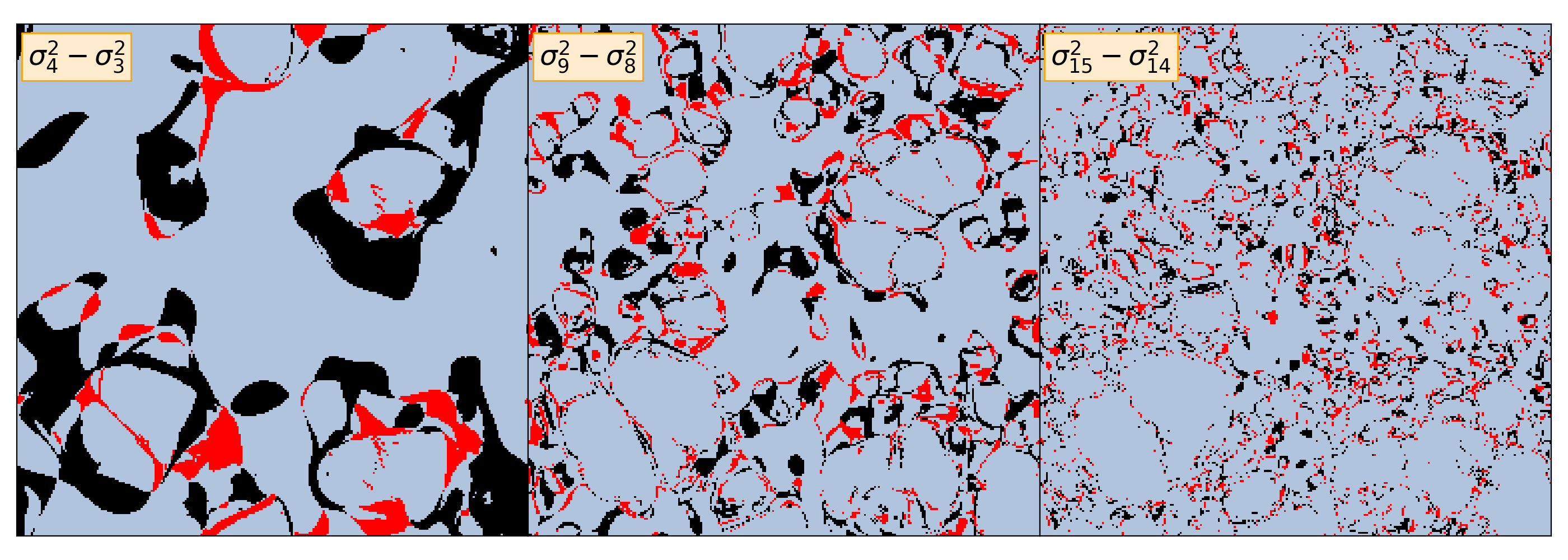} \\
    \includegraphics[width=1\textwidth]{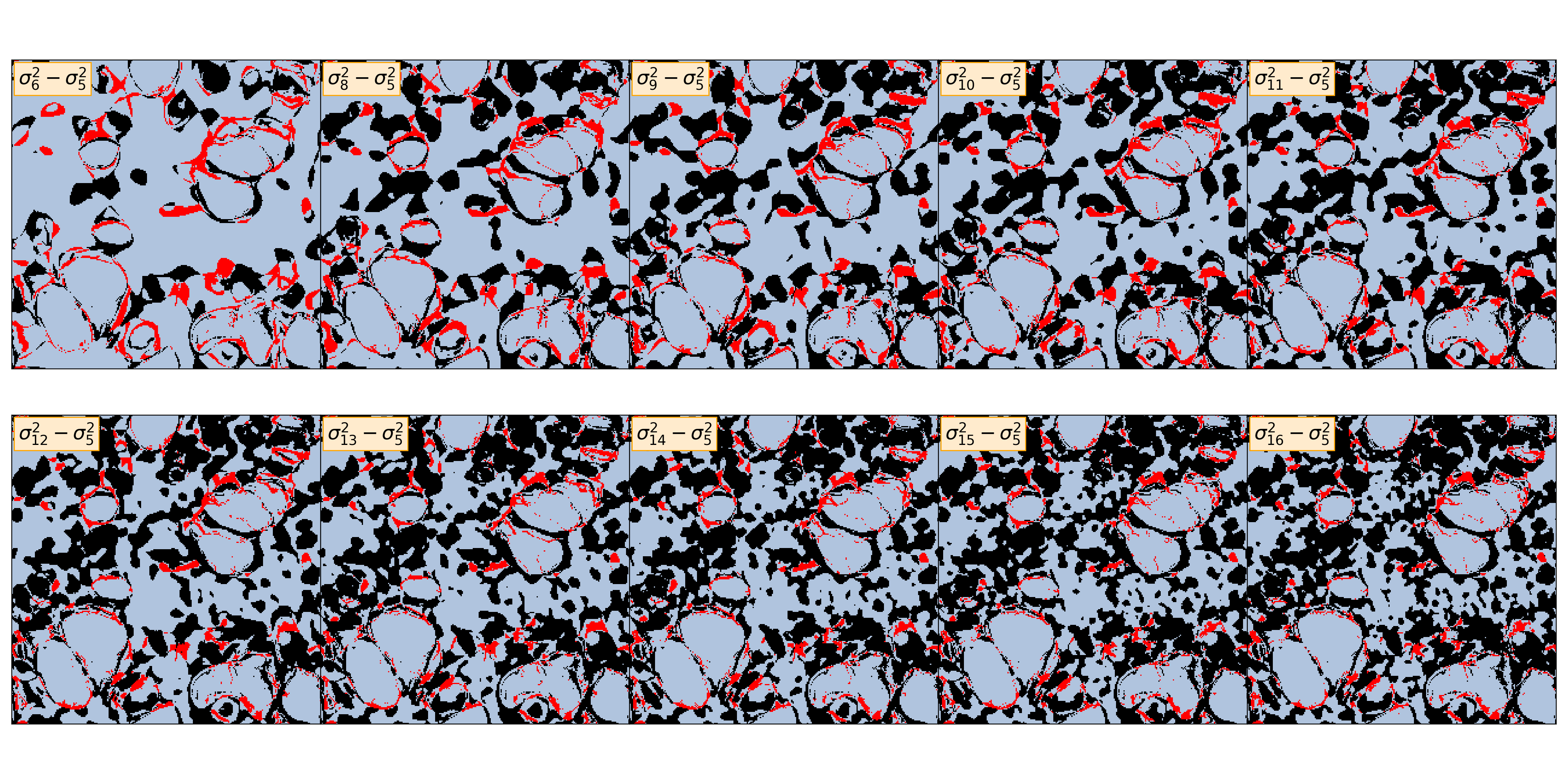} 
    \caption{Difference plots of $\Delta \sigma^2 = \sigma^2_i - \sigma^2_j$, where $i>j$. \textbf{Top:} Three examples of $\Delta \sigma^2 = 1$, where the scale of un-collapsing is most prominent. \textbf{Bottom:} Evolution of the difference field with increasing $\Delta \sigma^2$ (rightwards direction) for $\sigma^2_5$.}
    \label{fig_app:uncoll}
\end{figure*}

Figure \ref{fig_app:uncoll} presents several examples in Lagrangian coordinates, where un-collapse (red regions) is observed for different values of $\Delta \sigma^2$ in simulations smoothed with the sharp $k-$space window function. This plot compliments Figure \ref{fig:conf_matrices}, where it can be seen that as the scale of smoothing decreases, so does the scale of un-collapsed objects with an increasing $\Delta \sigma^2$. The gray regions represent coordinates where there is no difference between $\sigma^2_i$ and $\sigma^2_j$, while the black patches indicate a higher rank of collapse in $\sigma^2_i$ with respect to $\sigma^2_j$.

\section{Mass fractions}
\label{app: mass_fractions}

\begin{figure}
	\includegraphics[width=1\columnwidth]{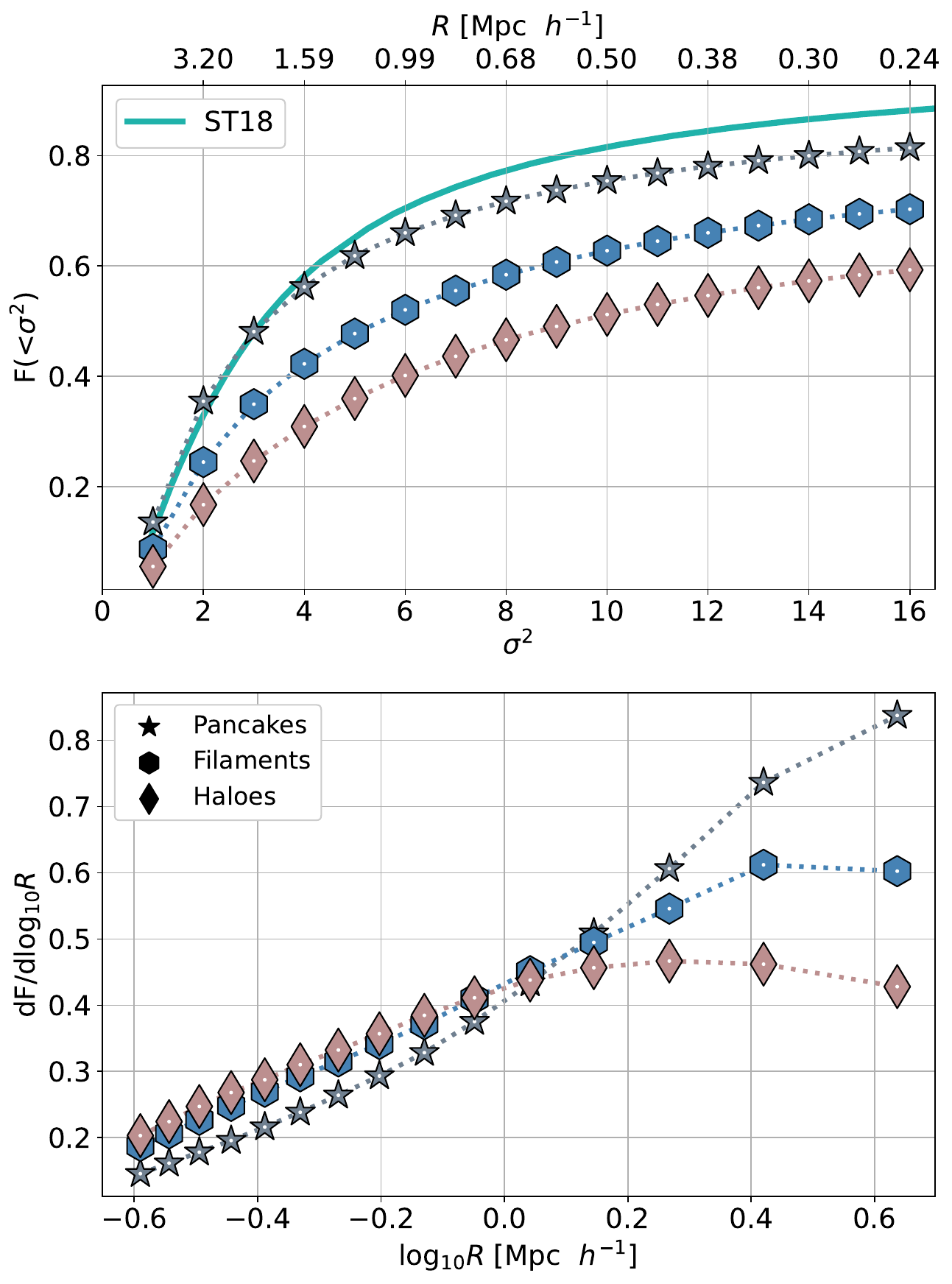}
    \caption{\textbf{Top:} The mass fractions of particles in pancakes (stars), filaments (hexagons) and haloes (diamonds) as a function of the smoothing scale $\sigma^2$ and $R$. The turquoise curve represents $F(<\sigma^2)$ of collapsed particles from \citet{stucker2018median}. \textbf{Bottom:} the particle fraction in structures of the ES radii. Both panels show the distributions for the $\widetilde{W}_{\tth}$ smoothed excursion set of simulations.}
    \label{fig:Mass_frract_th}
\end{figure}

Here we also present the mass fraction $F( < \sigma^2)$ plot for the example of the top-hat smoothing function. Notice how the turquoise curve, adapted from \citet{stucker2018median}, is not as good a match for the the mass fraction of $n \geq 1$ regions, as it was when the smoothing was carried out with the sharp$-k$ space filter. The reason for the discrepancy observed in Figure \ref{fig:Mass_frract_th}, is due to the fact that \citet{stucker2018median} also utilised the sharp$-k$ space filter for generating their random walks. Therefore, their results are more relevant for simulations smoothed out with this particular filter.

%%%%%%%%%%%%%%%%%%%%%%%%%%%%%%%%%%%%%%%%%%%%%%%%%%

% Don't change these lines
\bsp	% typesetting comment
\label{lastpage}
\end{document}